\documentclass[aps,prl,twocolumn,superscriptaddress,footinbib,linenumbers]{revtex4}
\usepackage{graphicx,bm,braket,exscale}
\usepackage[intlimits]{amsmath}
\usepackage{amsfonts}
\usepackage{amssymb,amscd}
\usepackage{color}
\usepackage{hyperref}
\usepackage{subfigure}
\usepackage[normalem]{ulem}
\usepackage{slashed}
\usepackage{leftidx}
\usepackage{xcolor,soul}

\soulregister\cite7
\soulregister\ref7
\sethlcolor{cyan}

\graphicspath{{Figures/}}

\newcommand{\Fig}[1]{Fig.~\ref{#1}}

\begin{document}


\title{Observation of gauge invariance in a 71-site Bose-Hubbard quantum simulator}

\author{Bing~Yang}
\email[Current address:]{Institut f\"ur Experimentalphysik, Universit\"at Innsbruck, Technikerstra{\ss}e 25, 6020 Innsbruck, Austria}
\affiliation{Hefei National Laboratory for Physical Sciences at Microscale and Department of Modern Physics, University of Science and Technology of China, Hefei, Anhui 230026, China}
\affiliation{Physikalisches Institut, Ruprecht-Karls-Universit\"{a}t Heidelberg, Im Neuenheimer Feld 226, 69120 Heidelberg, Germany}
\affiliation{CAS Centre for Excellence and Synergetic Innovation Centre in Quantum Information and Quantum Physics, University of Science and Technology of China, Hefei, Anhui 230026, China}

\author{Hui~Sun}
\affiliation{Hefei National Laboratory for Physical Sciences at Microscale and Department of Modern Physics, University of Science and Technology of China, Hefei, Anhui 230026, China}
\affiliation{Physikalisches Institut, Ruprecht-Karls-Universit\"{a}t Heidelberg, Im Neuenheimer Feld 226, 69120 Heidelberg, Germany}
\affiliation{CAS Centre for Excellence and Synergetic Innovation Centre in Quantum Information and Quantum Physics, University of Science and Technology of China, Hefei, Anhui 230026, China}

\author{Robert~Ott}
\affiliation{Institute for Theoretical Physics, Ruprecht-Karls-Universit\"{a}t Heidelberg, Philosophenweg 16, 69120 Heidelberg, Germany}

\author{Han-Yi~Wang}
\affiliation{Hefei National Laboratory for Physical Sciences at Microscale and Department of Modern Physics, University of Science and Technology of China, Hefei, Anhui 230026, China}
\affiliation{Physikalisches Institut, Ruprecht-Karls-Universit\"{a}t Heidelberg, Im Neuenheimer Feld 226, 69120 Heidelberg, Germany}
\affiliation{CAS Centre for Excellence and Synergetic Innovation Centre in Quantum Information and Quantum Physics, University of Science and Technology of China, Hefei, Anhui 230026, China}

\author{Torsten V.~Zache}
\affiliation{Institute for Theoretical Physics, Ruprecht-Karls-Universit\"{a}t Heidelberg, Philosophenweg 16, 69120 Heidelberg, Germany}

\author{Jad C.~Halimeh}
\affiliation{Department of Physics, University of Trento, Via Sommarive 14, 38123 Povo (TN), Italy}
\affiliation{Kirchhoff-Institute for Physics, Ruprecht-Karls-Universit\"{a}t Heidelberg, Im Neuenheimer Feld 227, 69120 Heidelberg, Germany}
\affiliation{Institute for Theoretical Physics, Ruprecht-Karls-Universit\"{a}t Heidelberg, Philosophenweg 16, 69120 Heidelberg, Germany}

\author{Zhen-Sheng~Yuan}
\email[e-mail:]{yuanzs@ustc.edu.cn}
\affiliation{Hefei National Laboratory for Physical Sciences at Microscale and Department of Modern Physics, University of Science and Technology of China, Hefei, Anhui 230026, China}
\affiliation{Physikalisches Institut, Ruprecht-Karls-Universit\"{a}t Heidelberg, Im Neuenheimer Feld 226, 69120 Heidelberg, Germany}
\affiliation{CAS Centre for Excellence and Synergetic Innovation Centre in Quantum Information and Quantum Physics, University of Science and Technology of China, Hefei, Anhui 230026, China}

\author{Philipp~Hauke}
\email[e-mail:]{philipp.hauke@unitn.it}
\affiliation{Department of Physics, University of Trento, Via Sommarive 14, 38123 Povo (TN), Italy}
\affiliation{Kirchhoff-Institute for Physics, Ruprecht-Karls-Universit\"{a}t Heidelberg, Im Neuenheimer Feld 227, 69120 Heidelberg, Germany}
\affiliation{Institute for Theoretical Physics, Ruprecht-Karls-Universit\"{a}t Heidelberg, Philosophenweg 16, 69120 Heidelberg, Germany}

\author{Jian-Wei~Pan}
\email[e-mail:]{pan@ustc.edu.cn}
\affiliation{Hefei National Laboratory for Physical Sciences at Microscale and Department of Modern Physics, University of Science and Technology of China, Hefei, Anhui 230026, China}
\affiliation{Physikalisches Institut, Ruprecht-Karls-Universit\"{a}t Heidelberg, Im Neuenheimer Feld 226, 69120 Heidelberg, Germany}
\affiliation{CAS Centre for Excellence and Synergetic Innovation Centre in Quantum Information and Quantum Physics, University of Science and Technology of China, Hefei, Anhui 230026, China}


\maketitle

\textbf{The modern description of elementary particles, as formulated in the Standard Model of particle physics, is built on gauge theories~\cite{Weinberg:2005}.
Gauge theories implement fundamental laws of physics by local symmetry constraints.
For example, in quantum electrodynamics, Gauss's law introduces an intrinsic local relation between charged matter and electromagnetic fields, which protects many salient physical properties including massless photons and a long-ranged Coulomb law.
Solving gauge theories by classical computers is an extremely arduous task~\cite{Gattringer:2009}, which has stimulated a vigorous effort to simulate gauge-theory dynamics in microscopically engineered quantum devices~\cite{Wiese:2013,Zohar:2015,Dalmonte:2016,Banuls:2019}.
Previous achievements implemented density-dependent Peierls phases without defining a local symmetry \cite{Clark:2018,Goerg:2019}, realized mappings onto effective models to integrate out either matter or electric fields \cite{Martinez:2016,Bernien:2017,Surace:2020,Kokail:2019}, or were limited to very small systems \cite{Dai:2017,Klco:2018,Schweizer:2019,Mil:2020}.
The essential gauge symmetry has not been observed experimentally.
Here, we report the quantum simulation of an extended U(1) lattice gauge theory, and experimentally quantify the gauge invariance in a many-body system comprising matter and gauge fields. These are realized in defect-free arrays of bosonic atoms in an optical superlattice of 71 sites.
We demonstrate full tunability of the model parameters and benchmark the matter--gauge interactions by sweeping across a quantum phase transition.
Enabled by high-fidelity manipulation techniques, we measure the degree to which Gauss's law is violated by extracting probabilities of locally gauge-invariant states from correlated atom occupations. Our work provides a way to explore gauge symmetry in the interplay of fundamental particles using controllable large-scale quantum simulators.}

\begin{figure}[!htb]
\includegraphics[width=7.2cm]{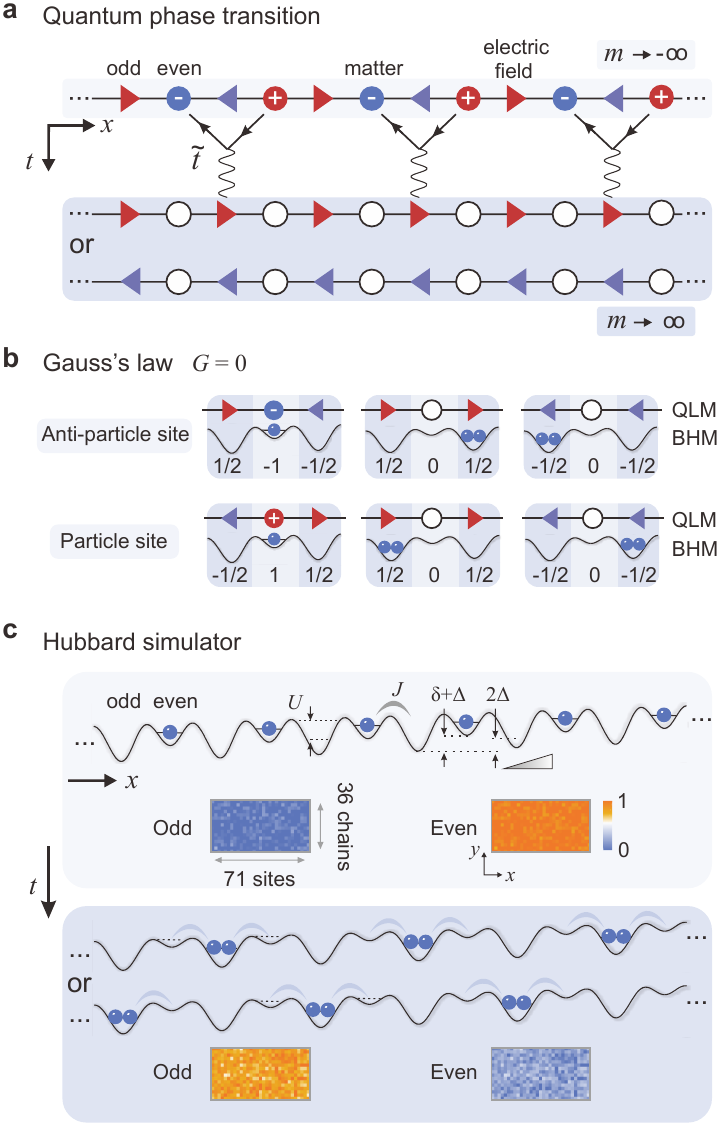}
\caption{Quantum simulation of a U(1) lattice gauge theory.
(a) A quantum phase transition separates a charge-proliferated phase from a C/P symmetry-breaking phase where the electric field (triangles) passes unhindered through the system (sketched at particle rest mass $m\to-\infty$ and $+\infty$, respectively). The transition is driven by the gauge-invariant annihilation of particles and antiparticles (charged circles).
(b) Gauss's law strongly restricts the permitted gauge-invariant configurations of charges and neighboring electric fields.
(c) Simulation of the model on a $71$-site Bose--Hubbard system consisting of ultracold atoms in an optical superlattice.
We sweep through the quantum phase transition by controlling the Hubbard parameters. Particle-antiparticle annihilation is realized by atoms initially residing on even (shallow) sites binding into doublons on odd (deep) sites (insets: atomic densities for initial and final states).
}
\label{Fig:f1}
\end{figure}

\begin{figure*}[!htb]
\includegraphics[width=17.1cm]{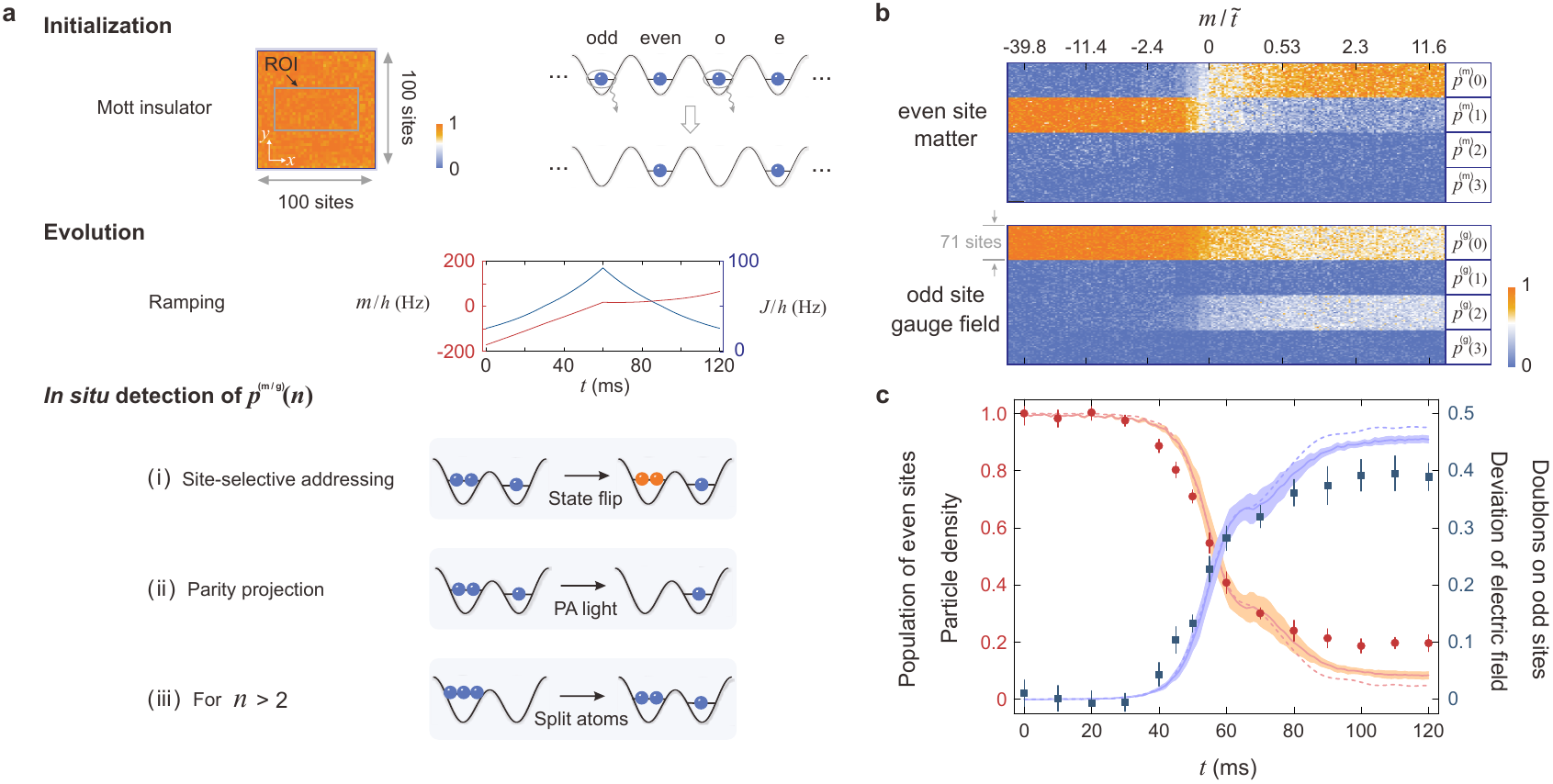}
\caption{
Probing the many-body dynamics.
(a) Experimental sequence. Starting from a near perfect Mott insulator in the ``short'' lattice, the initial staggered state is prepared by removing the atoms on odd sites. We drive the phase transition by ramping the mass $m = \delta - U/2$ and the tunneling $J$. Afterwards, the occupation probabilities $p^{(\mathrm{m/g})}(n)$ are identified for even and odd sites by engineering the atomic states with measurement schemes (i--iii).
(b-c) Time-resolved observation of the C/P-breaking phase transition.
As revealed by the probabilities, atoms initially residing on even sites bind into doublons on odd sites, corresponding to an annihilation of particles and a deviation of the electric field, quantified by $\sum_\ell (-1)^{\ell} \langle\hat{E}_{\ell,\ell+1}(t) - \hat{E}_{\ell,\ell+1}(0)\rangle/(2N)$.
Measured results agree well with theoretical predictions (solid curves) from the time-adaptive density matrix renormalization group ($t$-DMRG) method, where our numerics takes into account spatial inhomogeneity and sampling over noisy experimental parameters (see Methods). Error bars and shaded regions represent standard deviations throughout this article.
The dashed lines represent the exact evolution of the ideal QLM (see Methods).
}
\label{Fig:f2}
\end{figure*}

Quantum electrodynamics (QED), the paradigm example of a gauge-invariant quantum field theory, has fundamentally shaped our understanding of modern physics. Gauge invariance in QED---described as a local U(1) symmetry of the Hamiltonian---ties electric fields $\vec{E}$ and charges $\rho$ to each other through Gauss's law, $\nabla \cdot \vec{E} = \rho$.
This basic principle of gauge invariance has stood model for the Standard Model of particle physics, including, e.g., quantum chromodynamics. However, despite impressive feats \cite{Calzetta:2008,Berges:2004}, it remains extremely difficult for classical computers to solve the dynamics of gauge theories~\cite{Wiese:2013,Zohar:2015,Dalmonte:2016,Banuls:2019}. Quantum simulation offers the tantalizing prospect of sidestepping this difficulty by microscopically engineering gauge-theory dynamics in table-top experiments, based on, e.g. trapped ions, superconducting qubits, and cold atoms \cite{Martinez:2016,Bernien:2017,Surace:2020,Kokail:2019,Dai:2017,Klco:2018,Schweizer:2019,Goerg:2019,Clark:2018,Mil:2020}. In the quest for experimentally realizing gauge-theory phenomena, a large quantum system is essential to mitigate finite-size effects irrelevant to the theory in the thermodynamic limit. Moreover, while Gauss's law in QED holds \textit{fundamentally}, it is merely \textit{approximate} when engineered in present cold-atom experiments keeping both fermionic matter and dynamical gauge fields explicitly \cite{Schweizer:2019,Mil:2020}. Thus, it is a crucial challenge to determine the reliability of gauge invariance in large-scale quantum simulators \cite{Halimeh:2020}.

Here, we verify Gauss's law in a many-body quantum simulator. To this end, we devise a mapping from a Bose--Hubbard model describing ultracold atoms in an optical superlattice to a U(1) lattice gauge theory with fermionic matter. We exploit the formalism of quantum link models (QLMs) \cite{Chandrasekharan:1997,Wiese:2013}, which incorporates salient features of QED, in particular Coleman's phase transition in one spatial dimension (1D) at topological angle $\theta=\pi$ \cite{Coleman:1976}. Here, gauge invariant 'matter-gauge field' interactions emerge through a suitable choice of Hubbard parameters, effectively penalizing unwanted processes. Experimentally, we prepare large arrays of atoms in high-fidelity staggered chains, realize the quantum phase transition by slowly ramping the lattice potentials, and observe the characteristic dynamics via probing site occupancies and density--density correlations. In our model, Gauss's law constrains boson occupations over sets of three adjacent sites in the optical lattice. By tracking the coherent evolution of the state in these elementary units, we detect the degree of local violation of Gauss's law.

Our target model is a U(1) gauge theory on a 1D spatial lattice with $N$ sites, described by the Hamiltonian (see Methods)
\begin{equation}
\label{eq:QLM_Hamiltonian}
\hat{H}_{\mathrm{QLM}} = \sum_{\ell}\left[-\frac{\textrm{i}\tilde{t}}{2}  \left( \hat{\psi}_{\ell} \hat{S}^{+}_{\ell,\ell+1} \hat{\psi}_{\ell+1} - \mathrm{H.c.}\right) + m  \hat{\psi}^{\dagger}_{\ell}\hat{\psi}_{\ell}\right] \; .
\end{equation}
Using the QLM formalism, the gauge field is represented by spin-1/2 operators on links connecting neighboring lattice sites, $\hat{E}_{\ell,\ell+1} \equiv (-1)^{\ell+1} \hat{S}^{z}_{\ell,\ell+1}$, corresponding to an electric field coarse-grained to two values (red and blue arrows in Fig.~\ref{Fig:f1}). Using staggered fermions \cite{Susskind:1977}, matter fields $\hat{\psi}_{\ell}$ represent particles and anti-particles on alternating sites, with alternating electric charge $\hat{Q}_{\ell} = (-1)^\ell\hat{\psi}^{\dagger}_{\ell}\hat{\psi}_{\ell}$. By tuning the fermion rest mass $m$, we can drive the system across a quantum phase transition from a charge-dominated disordered phase to an ordered phase, characterized by the spontaneous breaking of charge and parity (C/P) symmetries \cite{Coleman:1976,Rico:2014}; see~\Fig{Fig:f1}a.
During the transition, due to the term $\propto \tilde{t}$, particle--anti-particle pairs annihilate accompanied by the correct adjustment of the electric field according to Gauss's law.

Gauss's law requires the generators of the U(1) gauge transformations,
\begin{align}
\hat{G}_\ell = (-1)^{\ell+1}\left(\hat{S}^z_{\ell,\ell+1} + \hat{S}^z_{\ell-1,\ell} + \hat{\psi}^{\dagger}_{\ell}\hat{\psi}_{\ell}\right) \; ,
\label{eq:Gauss-law}
\end{align}
to be conserved quantities for each matter site $\ell$. We choose, as is usual, to work in the charge-neutral sector, where the state $\ket{\psi}$ fulfills $\sum_{\ell} \hat{Q}_{\ell}\ket{\psi} = 0$, and in the Gauss's law sector specified by $\hat{G}_{\ell}\ket{\psi}=0$, $\forall \ell$. Ensuring adherence to this local conservation law is the main experimental challenge, as it intrinsically constrains matter and electric fields across three neighboring sites (see \Fig{Fig:f1}b).

We simulate this QLM with ultracold bosons in a 1D optical superlattice as sketched in \Fig{Fig:f1}c (see Methods for details). The experiment is governed by the Bose--Hubbard model (BHM)
\begin{equation}
\label{eq:BH_Hamiltonian}
\hat{H}_{\mathrm{BHM}}  = \sum_{j}\left[- J  (\hat{b}_j^{\dagger} \hat{b}_{j+1} + \mathrm{H.c.}) +  \frac{U}{2}\hat{n}_j (\hat{n}_j-1)  +   \varepsilon_j \hat{n}_j\right] ,
\end{equation}
where $\hat{b}_j^\dagger, \hat{b}_j$ are creation and annihilation operators, $\hat{n}_j=\hat{b}_j^\dagger \hat{b}_j$, and $J$ is the tunneling strength.
The energy offset $\varepsilon_j = (-1)^j\; \delta/2 + j\Delta$ consists of a linear tilt $\Delta$ to suppress long-range tunneling along the 1D chain and a staggered superlattice potential $\delta$. Here, the even sites $j$ of the superlattice correspond to the matter sites $\ell$ in the lattice gauge theory, while we identify odd sites $j$ with link indices $\ell,\ell+1$. Choosing $\delta\gg J$ and on-site interaction $U\sim 2\delta$, effectively constrains the system to the relevant subspace limited to the number states $\ket{0}$, $\ket{2}$ on odd (gauge) sites and $\ket{0}$, $\ket{1}$ on even (matter) sites.
On this subspace, we can hence identify $\hat{S}^+_{\ell,\ell+1}$$\simeq$$(\hat{b}_{j=2\ell+1}^{\dagger})^2/\sqrt{2}$ and (using a Jordan--Wigner transformation for the matter sites), and similarly $\hat{\psi}_{\ell} \hat{S}^{+}_{\ell,\ell+1} \hat{\psi}_{\ell+1}$ $\simeq$ $\hat{b}_{2\ell} (\hat{b}_{2\ell+1}^{\dagger})^2 \hat{b}_{2\ell+2}/\sqrt{2}$, see Methods.
This term can be physically realized by atoms on neighboring matter sites combining into a doublon.
The rest mass corresponds to $m = \delta - U/2$, which enables us to cross the phase transition by tuning $m <0 \rightarrow m>0$.
The strength of the gauge-invariant coupling ($\tilde{t} \approx 8\sqrt{2} J^2/U\approx 70$Hz at resonance $m \approx 0$) is much larger than the dissipation rate, enabling a faithful implementation in a large many-body system.

The experiment starts with a quasi two-dimensional Bose-Einstein condensate of about one hundred thousand $^{87}$Rb atoms in the $xy$-plane. We implement a recently demonstrated cooling method in optical lattices to create a Mott insulator with a filling factor of $0.992(1)$ \cite{Yang:2019}. Figure~\ref{Fig:f2}a shows a uniform area containing ten-thousand lattice sites, from which a region of interest (ROI) with $71\times36$ sites is selected for simulating the gauge theory. A lattice along the $y$-axis with depth $61.5(4)$ $E_r$ isolates the system into copies of 1D chains. Here, $E_r = h^2/(2 m_{\text{Rb}} \lambda_s^2)$ is the recoil energy, with $\lambda_s = 767$ nm the wavelength of a ``short'' lattice laser, $h$ the Planck constant, and $m_{\text{Rb}}$ the atomic mass. The near-unity filling enables the average length of defect-free chains to be longer than the $71$ sites. Even without a quantum gas microscope, the size of our many-body system is confirmed by counting the lattice sites with a single-site resonance imaging (see Methods). Along the $x$-direction, another ``long'' lattice with wavelength $\lambda_l=2\lambda_s$ is employed to construct a superlattice that divides the trapped atoms into odd and even sites. Two different configurations of the superlattice are used here. First, to manipulate quantum states in isolated double wells (DWs), which we use for state initialization and readout, the superlattice phase is controlled to match the positions of the intensity maxima of the short and long lattices. Second, in contrast, when performing the phase transition, overlapping the intensity minima of the lattices enables identical tunneling strength between neighboring sites.

To prepare the initial state, we selectively address and flip the hyperfine state of the atoms residing on odd sites \cite{Yang:2019}, followed by their removal using resonant light. The remaining atoms on the even sites of the 1D chains correspond to an overall charge neutral configuration. They form the ground state of our target gauge theory, Eq.~\eqref{eq:QLM_Hamiltonian}, at $m\rightarrow -\infty$ in the $\hat{G}_{\ell}\ket{\psi}=0$ sub-sector.

The phase transition is accessed by slowly tuning the superlattice structure in terms of the Hubbard parameters. The linear potential $\Delta = 57$ Hz/site (formed by the projection of gravity) as well as the main contribution to the staggered potential $\delta = 0.73(1)$ kHz (arising from the depth of the ``long'' lattice) are kept constant during the $120$ ms transition process. This ramp speed has been chosen to minimize both nonadiabatic excitations when crossing the phase transition and undesired heating effects. As shown in Fig.~\ref{Fig:f2}a, the tunneling strength $J/U$ is ramped from $0.014$ up to $0.065$ and back to $0.019$. Simultaneously, we linearly lower the $z$-lattice potential to ramp the on-site interaction $U$ from $1.82(1)$ kHz to $1.35(1)$ kHz. This ramp corresponds to driving the system from a large and negative $m$, through its critical point at $m \sim 0$, to a large and positive value deep within the C/P-broken phase.

\begin{figure}[!tb]
\includegraphics[width=8cm]{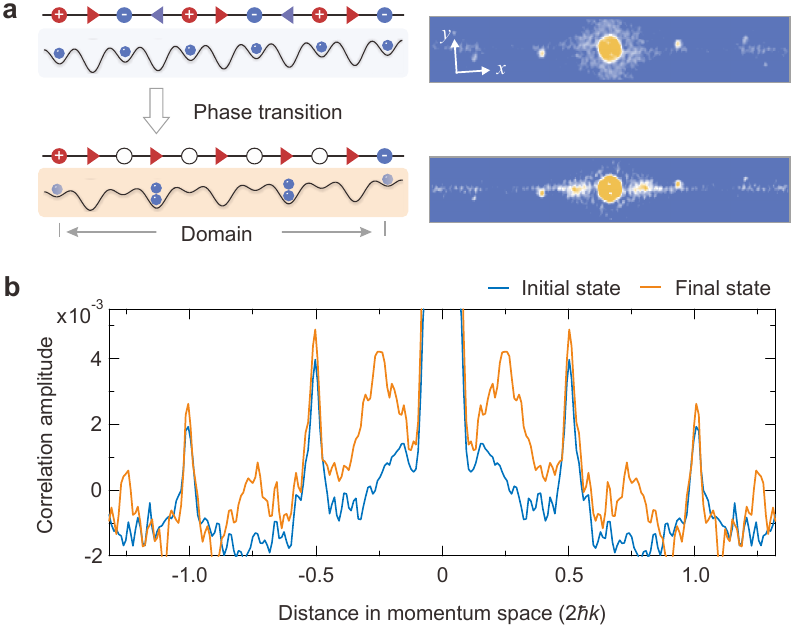}
\caption{Density-density correlation.
(a) Left: Idealized sketches of the initial and final state. The domain length of the final state equals to the distance between two unconverted atoms, which are removed from the system before measurement.
Right: Measured interference patterns in initial and final state (averaged over 523 and 1729 images, respectively).
The $x$-lattice defining the 1D chains is tilted by 4$^{\circ}$ relative to the imaging plane.
(b) Single-pixel sections along the $x$ direction through the center of the patterns in a. In the final state, additional peaks at $\pm 0.5\hbar k$ appear, indicating the emergence of a new ordering.
}
\label{Fig:f3}
\end{figure}

To probe the system dynamics, we ramp up the lattice barriers after evolution time $t$ and extract the probability distributions $p_j^{(\mathrm{m/g})}(n)$ of the occupation number $n$. With our optical resolution of $\sim$1 $\mu$m, \emph{in situ} observables average the signal over a small region around site $j$. Our measurements distinguish between even matter sites (m) and odd gauge-field sites (g). We illustrate the procedure for $p_j^{(\mathrm{g})}(n)$. To extract it for $n \le 3$, we combine the three schemes sketched in Fig.~\ref{Fig:f2}a (see Methods for a detailed translation of (i)-(iii) to the $p_j^{(\mathrm{m}/\mathrm{g})}(n)$). (i) The mean occupation of gauge-field sites is recorded by \emph{in situ} absorption imaging after applying a site-selective spin flip in the superlattice, which gives $\bar{n}^{(\mathrm{g})} = \sum_{n} n p_j^{(\mathrm{g})}(n)$ with natural numbers $n$. (ii) We use a photoassociation (PA) laser to project the occupancy into odd or even parity. Unlike selecting out doublons via Feshbach resonances \cite{Winkler:2006,Jordens:2008}, the PA-excited molecule decays spontaneously and gains kinetic energy to escape from the trap, with which the residual atomic density is $\bar{n}^{(\mathrm{g})}_c = \sum_{n} \text{mod}_2(n) p_j^{(\mathrm{g})}(n)$. (iii) A further engineering of atoms in DWs allows us to measure the probabilities of occupancies larger than two. We first clean the matter sites and then split the atoms into DWs. After a subsequent parity projection via illumination with PA light, the remaining atomic density is $\bar{n}^{(\mathrm{g})}_{c} + 2p_j^{(\mathrm{g})}(2)$. From the population, we find that high-energy excitations, such as $n = 3$, are negligible throughout our experiment.

\begin{figure}
\includegraphics[width=7.9cm]{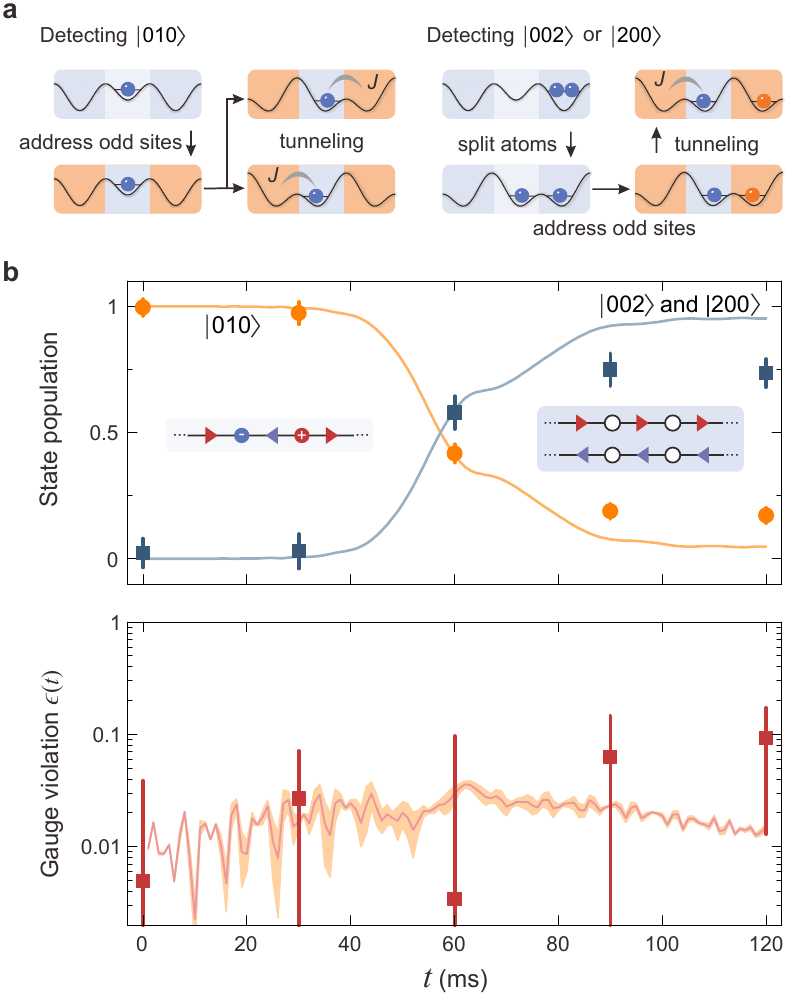}
\caption{Fulfillment of Gauss's law.
(a) Correlated measurements detect gauge-invariant states $\ket{...n_{j-1} n_{j} n_{j+1}...}$, $j$ even, within gauge-matter-gauge three-site units. For probing $\ket{...010...}$, we first flip the hyperfine levels of the atoms on odd sites. Then, we change the superlattice into two kinds of DW structures and monitor the tunneling of the middle atoms.
For $\ket{...002...}$ and $\ket{...200...}$, we split the doublons into two sites and mark them by the hyperfine levels. Their state populations correlate to the oscillation amplitudes of tunneling dynamics.
(b) From the probabilities of the gauge-invariant states, we extract the gauge violation $\epsilon(t) = 1- \left(p_{\ket{...010...}} + p_{\ket{...002...}} +p_{\ket{...200...}} \right)$.
While the inversion between the Fock states after the phase transition is stronger in the ideal QLM (exact numerics, orange and blue curves),
a high level of gauge invariance is retained throughout. The experimental results are in quantitative agreement with $t$-DMRG calculations for our isolated Bose--Hubbard system (red curve).
}
\label{Fig:f4}
\end{figure}

As the data for $p_j^{(\mathrm{m}/\mathrm{g})}(n)$ in Fig.~\ref{Fig:f2}b-c shows, after the ramp through the phase transition, on average $80(3)\%$ of the atoms have left the even sites and $39(2)\%$ of double occupancy is observed on the odd sites (we checked the coherence and reversibility of the process by ramping back from the final state, see Methods). This corresponds to the annihilation of $78(5)\%$ of particle--anti-particle pairs. From the remaining $22(5)\%$ of particles that have not annihilated, we estimate the average size of ordered domains after the ramp to be $9(2)$ sites. The formation of ordered domains can be further confirmed by measuring density-density correlations $C(i,j)=\langle \hat{n}_i \hat{n}_{j} \rangle$ \cite{Altman:2004,Folling:2005,Simon:2011}. We extract the correlation functions in momentum space after $8$ ms time of flight. For a bosonic Mott state with unity filling, the correlation function shows a bunching effect at momentum positions of $\pm 2\hbar k$, where $k=2\pi/\lambda_s$ is the wave vector. Two more peaks at $\pm \hbar k$ appear in the correlation function of our initial state due to the staggered distribution, as shown in Fig.~\ref{Fig:f3}a. The width of these peaks is mainly determined by the spatial resolution of the absorption imaging. The correlation function of the final state in Fig.~\ref{Fig:f3} shows two broader peaks at $\pm 0.5 \hbar k$, which indicates the emergence of a new ordering with a doubled spatial period. The finite correlation length $\xi$ of the final state broadens the interference pattern. Assuming exponential decay of density-density correlations, $C(i,j) \propto \exp(-|i-j|/\xi)$, we obtain the correlation length of the final state to be $\xi = 4.4^{+2.0}_{-1.0}$ sites (see Methods). Thus, we can achieve many-body regions with spontaneously broken C/P symmetry.

Finally, we quantify the violation of Gauss's law, for which we monitor the probabilities $p_{\ket{...n_{j-1} n_{j} n_{j+1}...}}$ of the three allowed gauge-invariant Fock states sketched in Fig.~\ref{Fig:f1}b, $\ket{...n_{j-1} n_{j} n_{j+1}...}=\ket{...010...}$, $\ket{...200...}$, and $\ket{...002...}$, $j$ even. To achieve this, we have developed a method to measure the density correlations between neighboring lattice sites within DWs. Unlike the approach in Fig.~\ref{Fig:f2}a, which does not give access to correlations between sites, here we distinguish different states by their dynamical features (Fig.~\ref{Fig:f4}a). In particular, we use the characteristic tunneling frequency to distinguish the target states from the others. For example, to detect the state $\ket{...010...}$, we perform tunneling sequences between DWs in two mirrored superlattice configurations (setting the parameters to $J/h = 68.9(5)$ Hz and $U/h = 1.71(1)$ kHz to avoid frequency overlap between different processes). The tunneling frequency $2J/h$ for the state $\ket{10}$ in a DW 
is one order of magnitude higher than the superexchange frequency $4J^2/hU$ for the states $\ket{20}$ or $\ket{11}$. Thus, the oscillation amplitudes at frequency $2J/h$ yield the probabilities $p_{\ket{...01n_{j+1}...}}$ and $p_{\ket{...n_{j-1}10...}}$. In addition, the probability $p_{\ket{...n_{j-1}1n_{j+1}...}}$ equals to $p_j^{\mathrm{(m)}}(1)$ (see Fig.~\ref{Fig:f2}b-c). With these, we can deduce a lower bound $p_{\ket{...010...}} \geq  p_{\ket{...01n_{j+1}...}} + p_{\ket{...n_{j-1}10...}} - p_{\ket{...n_{j-1}1n_{j+1}...}}$.
We obtain the population of the states $\ket{...002...}$ and $\ket{...200...}$ in a similar fashion (see Methods).

From these measurements, we can obtain the degree of gauge violation $\epsilon(t)$, defined as the spatial average of $1- \bra{\psi(t)}P_{\ell}\ket{\psi(t)}$, where $P_{\ell}$ projects the system state $\ket{\psi(t)}$ onto the local gauge-invariant subspace. As shown in Fig.~\ref{Fig:f4}b, throughout our entire experiment the summed probabilities of gauge-invariant states remains close to 1. Thus, our many-body quantum simulator retains gauge invariance to an excellent degree, even during and after a sweep through a quantum phase transition.

\
\

In conclusion, we have developed a fully tunable many-body quantum simulator for a U(1) gauge theory and demonstrated that it faithfully implements gauge invariance, the essential property of lattice gauge theories. Future extensions may give access to other symmetry groups and gauge theories in higher dimensions. The main challenge for the latter is to combine the model with a plaquette term that has been demonstrated previously in the present apparatus \cite{Dai:2017}. Importantly, our results enable the controlled analysis of gauge theories far from equilibrium, which is notoriously difficult for classical computers~\cite{Wiese:2013,Zohar:2015,Dalmonte:2016,Banuls:2019}. A plethora of fascinating target phenomena offers itself for investigation, including false vacuum decay \cite{Hauke:2013,Yang:2016}, dynamical transitions related to the topological $\theta$-angle \cite{Zache:2019,Huang:2019,Magnifico:2019}, and thermal signatures of gauge theories under extreme conditions \cite{Berges:2018}.



\clearpage
\newpage
\onecolumngrid
\vspace*{0.5cm}
\begin{center}
    \textbf{METHODS}
\end{center}
\vspace*{0.5cm}
\twocolumngrid
\newpage

\setcounter{figure}{0}
\makeatletter
\makeatother
\renewcommand{\figurename}{Extended Data Fig.}

\subsection{Target Model}

Our experiment is motivated by the lattice Schwinger model of QED in one spatial dimension in a Kogut--Susskind Hamiltonian formulation \cite{Kogut:1975s},
\begin{align}
\label{eq:supp_theory1}
\hat{H}_{\mathrm{QED}} = \frac{a}{2}& \sum_{\ell} \left( \hat{E}_{\ell,\ell+1}^2 +m (-1)^{\ell} \hat{\psi}^{\dagger}_{\ell}\hat{\psi}_{\ell} \right) \nonumber\\
&-\frac{\text{i}}{2a}\sum_{\ell} \left(\hat{\psi}_{\ell}^{\dagger} \hat{U}_{\ell,\ell+1} \hat{\psi}_{\ell+1} - \mathrm{H.c.} \right)  \; ,
\end{align}
with lattice spacing $a$, gauge coupling $e$, and where we have set $\hbar$ and $c$ to unity for notational brevity. Gauge links and electric fields fulfill the commutation relations $[\hat{E}_{\ell,\ell+1},\hat{U}_{m,m+1}] = e \delta_{\ell,m}\hat{U}_{\ell,\ell+1}$, while fermion field operators obey canonical anti-commutation relations $\{\hat{\psi}_{\ell}^{\dagger},\hat{\psi}_m\} = \delta_{\ell m}$. Here, we use `staggered fermions' \cite{Susskind:1977}, which are an elegant way to represent oppositely charged particles and antiparticles, using a single set of spin-less fermionic operators, but at the expense of alternating signs on even and odd sites.

Gauge transformations are expressed in terms of the local Gauss's law operators
\begin{equation}
\hat{G}_{\ell} = \hat{E}_{\ell,\ell+1} - \hat{E}_{\ell-1,\ell} - e \frac{\hat{\psi}^{\dagger}_{\ell} \hat{\psi}_\ell + (-1)^{\ell}}{2} \; .
\end{equation}
These generate local U(1) transformations parametrized by real numbers $\alpha_\ell$, under which an operator $\hat{\mathcal{O}}$ transforms as $\hat{\mathcal{O}}' =  \hat{V}^\dagger \hat{\mathcal{O}} \hat{V}$, with $\hat{V} = \exp \left[i \sum_\ell \alpha_\ell \hat{G}_\ell\right]$. Explicitly, the matter and gauge fields transform according to $\hat{\psi}_\ell' = e^{-ie\alpha_\ell} \hat{\psi}_\ell$, $\hat{U}'_{\ell,\ell+1} =  e^{-ie\alpha_\ell}  \hat{U}_{\ell,\ell+1} e^{ie\alpha_{\ell+1}}$ and $\hat{E}_{\ell,\ell+1}' = \hat{E}_{\ell,\ell+1}$.
In the absence of external charges, a physical state $\ket{\psi(t)}$ is required to be invariant under a gauge transformation, i.e. $\hat{V} \ket{\psi(t)} = \ket{\psi(t)}$.
Thus, gauge invariance under Hamiltonian time evolution is equivalent to $\hat{G}_{\ell}\ket{\psi(t)} = 0$, $\forall \ell, t$, i.e., $[\hat{G}_{\ell},\hat{H}_{\mathrm{QED}}] = 0$ and the $\hat{G}_\ell$ are conserved charges. In our experiments, we achieve to explicitly probe this local conservation law (see Fig.~\ref{Fig:f4}b and further below).

\begin{figure}[!tb]
\includegraphics[width=8.2cm]{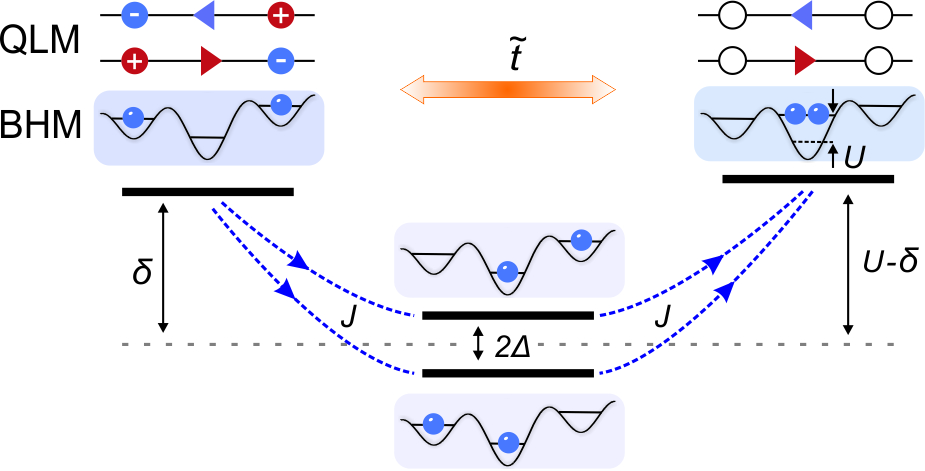}
\caption{Level structure of a three-site building block (matter-gauge-matter). The energy manifold of interest is given by the state on the left representing a particle pair, and the one on the right, where particles have annihilated while changing the in-between gauge field configuration. In the middle, we show the detuned intermediate processes and states by which these ``physical'', i.e., gauge-invariant states are coupled.}\label{Fig:Supp_theory1}
\end{figure}

Using the QLM formalism \cite{Chandrasekharan:1997}, we represent U(1) gauge fields in our analog quantum simulator by spin-$1/2$ operators $\hat{U}_{\ell,\ell+1} \rightarrow \frac{2}{\sqrt{3}}\hat{S}^+_{\ell,\ell+1}$ ($\frac{2}{\sqrt{3}}\hat{S}^-_{\ell,\ell+1}$) for even (odd) $\ell$, as well as $\hat{E}_{\ell,\ell+1} \rightarrow e(-1)^{\ell}\hat{S}^z_{\ell,\ell+1}$. In this spin-$1/2$ QLM representation, the electric field energy term $\sim \hat{E}^2$ represents a constant energy offset, and hence drops out. Therefore, without loss of generality, we may set $e \rightarrow 1$ in the following. With the rather untypical sign conventions in the above QLM definition, and an additional particle-hole transformation on every second matter site ($\hat{\psi}_\ell\leftrightarrow \hat{\psi}_\ell^\dagger$, $\ell$ odd), the alternating signs of the staggered fermions are canceled, yielding a simpler homogeneous model \cite{Hauke:2013,Yang:2016}. The Hamiltonian takes the form \eqref{eq:QLM_Hamiltonian} and Gauss's law is represented by Eq.~\eqref{eq:Gauss-law}.

For large \emph{negative} values of the mass, $m\rightarrow -\infty$, the ground state is given by fully occupied fermion sites and an alternating electric field $\hat{E}$.
However, for large \emph{positive} masses, the absence of fermions is energetically favorable. In this configuration there are no charges, and hence the electric fields are aligned, in a superposition of all pointing to the left and all pointing to the right (see Fig.~\ref{Fig:f1}). In between these two extreme cases, the system hosts a second-order quantum phase transition, commonly termed Coleman's phase transition \cite{Coleman:1976,Pichler:2016s}. While the quantum link Hamiltonian \eqref{eq:QLM_Hamiltonian} is invariant under a transformation \cite{Pichler:2016s} of parity (P) and charge conjugation (C) the ground state does not always respect these symmetries: The vacuum state for $m\rightarrow -\infty$ is C- and P-invariant, but the respective vacua in the $m\rightarrow \infty$ phase are C- and P-broken. An order parameter for the transition is given by the staggered change of the electric fields with respect to the initial configuration, $\sum_{\ell} \langle\hat{E}_{\ell,\ell+1}(t)- \hat{E}_{\ell,\ell+1}(0)\rangle/(2N)$.

\subsection{Mapping to Bose--Hubbard simulator}

Starting from the Hamiltonian \ref{eq:QLM_Hamiltonian}, we employ a Jordan--Wigner transformation with alternating minus signs,
\begin{subequations}
\begin{align}
\label{eq:JW}
\hat{\psi}_\ell^{\dagger} &= (-1)^{\ell} e^{i\pi \sum_{\ell'=0}^{\ell-1} ((-1)^{\ell'}\hat{\sigma}^{z}_{\ell'}+1)/2 } \hat{\sigma}_\ell^{+} \; ,\\
\hat{\psi}_\ell &= (-1)^{\ell} e^{-i\pi \sum_{\ell'=0}^{\ell-1} ((-1)^{\ell'}\hat{\sigma}^{z}_{\ell'}+1)/2 } \hat{\sigma}_\ell^{-} \; ,\\
\hat{\psi}^{\dagger}_\ell & \hat{\psi}_\ell = \frac{\hat{\sigma}^{z}_\ell+1}{2} \; ,
\end{align}
\end{subequations}
replacing the fermionic operators $\hat{\psi}_{\ell}$/$\hat{\psi}^{\dagger}_{\ell}$ by local spin-$1/2$ operators $\hat{\sigma}^{\pm}_{\ell}$ and non-local strings involving $\hat{\sigma}^{z}_{\ell'<\ell}$. We further identify the eigenstates of $\hat{\sigma}^z_\ell$ with two bosonic harmonic oscillator eigenstates $\ket{0}_{\ell}$ and $\ket{1}_{\ell}$. Projecting to the subspace $\mathcal{H}_{\ell} = \text{span}\{\ket{0}_\ell,\ket{1}_\ell\}$, we then realize the spin operators in terms of bosonic creation/annihilation operators $\hat{a}^{\dagger}$/$\hat{a}$ as follows:

\begin{subequations}
\begin{align}
	\hat{\sigma}_\ell^{-} &= \mathcal{P}_\ell \hat{a}_\ell \mathcal{P}_\ell\; ,\\
	\hat{\sigma}_\ell^{+} &= \mathcal{P}_\ell\hat{a}^{\dagger}_\ell \mathcal{P}_\ell\; ,\\
	\hat{\sigma}^z_\ell &= \mathcal{P}_\ell\left(2\hat{a}_\ell^{\dagger}\hat{a}_\ell - 1\right)\mathcal{P}_\ell  \; ,
\end{align}
\end{subequations}
where $\mathcal{P}_\ell$ is the projector onto $\mathcal{H}_\ell$. The bosonic commutation relations $[\hat{a}_\ell,\hat{a}^{\dagger}_\ell] = 1$ when restricted to $\mathcal{H}_\ell$ imply the required algebra of the Pauli matrices as given by $[\hat{\sigma}^{z}_\ell,\hat{\sigma}^{\pm}_\ell] = \pm 2\hat{\sigma}^{\pm}_\ell$ and $[\hat{\sigma}^{+}_\ell,\hat{\sigma}^{-}_\ell] =  \hat{\sigma}^{z}_\ell$. Similarly, we identify the two eigenstates of the "gauge" spins $\hat{S}^z_{\ell,\ell+1}$ with two eigenstates $\ket{0}_{\ell,\ell+1}$ and $\ket{2}_{\ell,\ell+1}$ associated with further bosonic operators, $\hat{d}_{\ell,\ell+1}$ and $\hat{d}_{\ell,\ell+1}^{\dagger}$ located at the links. Projecting to the subspace $\mathcal{H}_{\ell,\ell+1} = \text{span}\{\ket{0}_{\ell,\ell+1},\ket{2}_{\ell,\ell+1}\}$, we have

\begin{subequations}
\begin{align}
\label{eq:S_minus}
\hat{S}^{-}_{\ell,\ell+1} &= \frac{1}{\sqrt{2}}\mathcal{P}_{\ell,\ell+1} \left(\hat{d}_{\ell,\ell+1}\right)^2 \mathcal{P}_{\ell,\ell+1}\; ,\\
\hat{S}^{+}_{\ell,\ell+1} &= \frac{1}{\sqrt{2}}\mathcal{P}_{\ell,\ell+1}\left(\hat{d}^{\dagger}_{\ell,\ell+1}\right)^2 \mathcal{P}_{\ell,\ell+1}\; ,\\
\hat{S}^{z}_{\ell,\ell+1} &= \frac{1}{2}\mathcal{P}_{\ell,\ell+1}\left(\hat{d}_{\ell,\ell+1}^{\dagger}\hat{d}_{\ell,\ell+1} - 1\right)\mathcal{P}_{\ell,\ell+1}  \; ,
\end{align}
\end{subequations}
fulfilling the desired angular momentum algebra. With these replacements, the Hamiltonian \ref{eq:QLM_Hamiltonian} becomes

\begin{align}
\label{eq:supp_theory3}
\hat{H}_{\mathrm{QLM}} =\mathcal{P} \sum_{\ell}\left\{ m \hat{a}^{\dagger}_{\ell} \hat{a}_{\ell} + \frac{\tilde{t}}{2\sqrt{2}} \big[ \hat{a}_{\ell} (\hat{d}^{\dagger}_{\ell,\ell+1})^2 \hat{a}_{\ell+1} + \mathrm{H.c.}\big]\right\}\mathcal{P}\; ,
\end{align}
where $\mathcal{P} = \prod_\ell \mathcal{P}_\ell \mathcal{P}_{\ell,\ell+1}$. In the main text, the projection $\mathcal{P}$ is implied in the notation $\hat{A}\simeq \hat{B}$, which abbreviates the equality $ \hat{A} =\mathcal{P}\hat{B}\mathcal{P}$ for two operators $\hat{A}$ and $\hat{B}$. We emphasize that even though Eq.~\eqref{eq:supp_theory3} is written in terms of bosonic operators, the projectors together with the Jordan--Wigner transform ensure--at the level of the Hamiltonian and diagonal observables--the equivalence with the original lattice gauge theory including fermionic matter.

The Hamiltonian in Eq.~\eqref{eq:supp_theory3} is generated effectively in our Bose--Hubbard system through a suitable tuning of the parameters described in Eq.~\eqref{eq:BH_Hamiltonian}.
As a preceding step, matter sites are identified with even sites of the optical superlattice ($\hat{a}_{\ell}\to \hat{b}_{j = 2\ell}$, $\ell= 0,..,N-1$) and gauge links with odd sites of the superlattice ($\hat{d}_{\ell,\ell+1}\to \hat{b}_{j = 2\ell+1}$, $\ell= 0,..,N-2$). For our quantum simulator, we have $N=36$ matter sites and $35$ gauge links, yielding a total of 71 bosonic sites. The principle for generating the gauge-invariant dynamics is then conveniently illustrated in a three-site building block consisting of optical-lattice sites $j$=0,1,2, as shown in Extended Data Fig. 1. The system is initialized in a state where all matter sites are singly occupied, while gauge links are empty, i.e., the system starts in the boson occupation state $\ket{101}$. By choosing $U,\delta \gg J$ and $U\sim 2\delta$, the two states $\ket{101}$ and $\ket{020}$ form an almost degenerate energy manifold $\alpha$ (not the absolute ground-state manifold of the BHM). This manifold is well-separated from the states $\ket{110}$ and $\ket{011}$, shown in the middle. Hence, direct tunneling of the bosons into (and out of) the deep gauge-link well is energetically off-resonant and suppressed. The effective dynamics between the states within the manifold $\alpha$ is then described by degenerate perturbation theory \cite{Lewenstein:2012s}, leading to the Hamiltonian \ref{eq:supp_theory3} acting on the subspace indicated by the projectors $\mathcal{P}$. An explicit calculation yields the effective coupling

\begin{equation}
\tilde{t} = \sqrt{2} J^2 \left(\frac{1}{\delta+\Delta} + \frac{1}{U-\delta +\Delta} + \frac{1}{\delta-\Delta} + \frac{1}{U-\delta-\Delta}\right),
\end{equation}
which reduces to a simple relation close to resonance, $\tilde{t} \overset{U\approx2\delta}{\rightarrow} 8\sqrt{2} J^2/U$. A key ingredient for this manner of generating the term $\propto \tilde{t}$ was the particle-hole transformation \cite{Hauke:2013,Yang:2016}. It enabled us to rewrite this term, which is usually interpreted as a kinetic hopping term, as the simultaneous motion of two bosons on neighboring matter sites into the gauge link in between (and back). Note that our approach of constraining to an energy manifold $\alpha$ from the total Hilbert space is different from previous works, where the authors proposed to implement gauge symmetry in the ground state manifold by adding a term $\sim \sum_x G_{x}^2$ to the Hamiltonian \cite{Zohar:2011, Banerjee:2013,Hauke:2013,Halimeh:2020}.

The above result, where we include couplings of our initial-state manifold to other manifolds at order $J$, is valid for both the building block as well as for the extended system close to resonance. In our many-body system, at this order in perturbation theory, the mass is represented by the energy imbalance of on-site interaction and staggering, $m = \delta - U/2$, such that the gauge-invariant particle creation/annihilation becomes resonant once fermions are massless. In the chosen parameter configuration, occupations other than the desired $\ket{0}$,$\ket{1}$ (even) and $\ket{0}$,$\ket{2}$ (odd sites) are highly suppressed, as we confirmed through numerics and direct measurement (see Fig.~\ref{Fig:f2}b). We also include a linear tilt potential to suppress the tunneling of atoms to their next-nearest neighboring sites, e.g., $\ket{02001} \longleftrightarrow \ket{02100}$ (here $j=0,...,4$), as such processes are also generated at second order and would break gauge invariance.

\subsection{State preparation and detection}

The experiment begins with a quasi two-dimensional quantum gas of $\sim$8.6$\times 10^4$ atoms prepared by adiabatically loading a nearly pure Bose--Einstein condensate into a single well of a pancake-shaped standing wave. The pancake trap is generated by interfering two blue-detuned laser beams at wavelength $\lambda_s = 767$ nm, which provides the confinement along the $z$-axis. We implement a staggered-immersion cooling for the quantum gas to create a Mott insulator with near-unity filling \cite{Yang:2019}. The cooling is performed within an optical superlattice where the atoms are separated into superfluid and Mott-insulator phases with a staggered structure. The superlattice potential can be written as
\begin{equation}
V(x)=V_s\mathrm{cos}^2(kx)-V_l\mathrm{cos} ^2(kx/2+\varphi).
\end{equation}
Here, the relative phase $\varphi$ determines the superlattice configuration, which is controlled by changing the relative frequency of these lasers. At $\varphi=0$, the atoms on odd and even sites of the double wells experience the same trap potential. In the cooling stage, we keep the phase at $\varphi = 7.5(7)$ mrad to generate a staggered energy difference for the odd and even sites. After cooling, the temperature of the Mott-insulator sample with $\overline{n}=2$ is $k_B T_f = 0.046(10) U$. Then, we freeze the atomic motions and remove the high-entropy atoms. Based on the low-entropy sample, a Mott insulator with $99.2(1) \%$ of single occupancy is prepared by separating the atom pairs within the double-well superlattice. Figure \ref{Fig:f2}a shows such a two-dimensional sample with a homogeneous regime containing $10^4$ lattice sites.

The technique of site-selective addressing is widely used in our experiment \cite{Yang:2017s,Yang:2019}. For the Mott insulator, all the atoms are prepared in the hyperfine level of $\ket{\downarrow}= \Ket{F=1, m_F=-1}$. We define another pseudo-spin state as $\ket{\uparrow}= \Ket{F=2, m_F=-2}$. When the direction of the bias field is along $x$ and the phase of the electro-optical modulator is set to $\pi/3$, the energy splitting between $\ket{\downarrow}$ and $\ket{\uparrow}$ has a $28$-kHz difference for the odd and even sites. We edit the microwave pulse and perform a rapid adiabatic passage to selectively flip the hyperfine states of atoms on odd or even sites, achieving an efficiency of $99.5(3)\%$. For state initialization, we flip the atomic levels of odd sites and then remove these atoms with a resonant laser pulse. This site-selective addressing is also employed for state readout, as shown in Fig.~\ref{Fig:f1}b.
Combining such technique with the absorption imaging, we record the atomic densities of odd and even sites successively in a single experimental sequence.

We use a parity projection of the atom number to probe the distribution of site occupancies. The basic idea is to remove the atom pairs by exciting them to a non-stable molecular state via the photoassociation (PA) process. The laser frequency is 13.6 $\text{cm}^{-1}$ red-detuned to the D2 line of $^{87}$Rb atom, which drives a transition to the $v$=17 vibrational state in the $0_g^-$ channel. The decay rate of the atom pairs is 5.6(2) kHz in the laser intensity of 0.67 W/cm$^{2}$. After applying this PA light for 20 ms, the recorded atom loss equals to the ratio of atom pairs. For detecting the filling number of more than double occupancy, we first engineer the atoms in the double wells and then detect the number parity with PA collision. As shown in Fig.~\ref{Fig:f2}a, we clear up the atoms occupying the even sites and then split the atoms of odd sites into double wells. If the occupancy is more than two, we can observe some atom loss after applying the PA light. The remaining atom number after this operation is $\bar{n}^{\mathrm{(g)}}_t = \bar{n}_c + 2 p^{\mathrm{(g)}}(2)$. From these measurements, we obtain the upper bound of the probabilities for these highly excited states, such as three or four atoms. Here, we consider the excitations up to three atoms with the probability $p^{\mathrm{(m/g)}}(3)$. Hence, the probabilities of matter or gauge sites derived from these detections are,
\begin{equation}
\begin{split}
p^{\mathrm{(m/g)}}(0)&=1- \frac{1}{2}\left[ \bar{n}^{\mathrm{(m/g)}}_c + \bar{n}^{\mathrm{(m/g)}}_t \right],\\
p^{\mathrm{(m/g)}}(1)&=\bar{n}^{\mathrm{(m/g)}}_c - \frac{1}{2}\left[\bar{n}^{\mathrm{(m/g)}} - \bar{n}^{\mathrm{(m/g)}}_t\right],\\
p^{\mathrm{(m/g)}}(2)&=\frac{1}{2}\left[\bar{n}^{\mathrm{(m/g)}}_t -\bar{n}^{\mathrm{(m/g)}}_c\right],\\
p^{\mathrm{(m/g)}}(3)&=\frac{1}{2}\left[\bar{n}^{\mathrm{(m/g)}}-\bar{n}^{\mathrm{(m/g)}}_t\right].
\end{split}
\end{equation}

These probabilities refer to the observables given by the detection methods (i-iii) in our main text.

\subsection{Imaging individual sites in the 1D chain without a quantum gas microscope}

\begin{figure}[!htb]
\includegraphics[width=8cm]{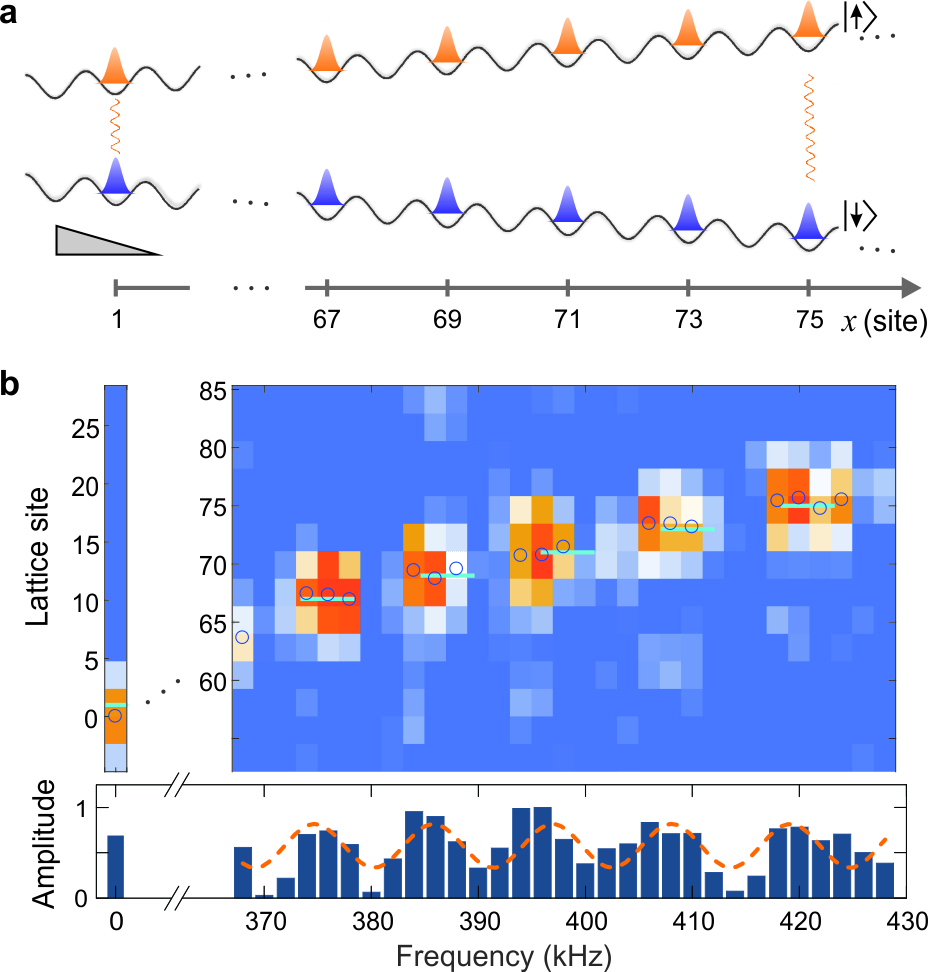}
\caption{Single-site resolved imaging. (a) We sketch the staggered filling in a 1D chain as our initial state for the quantum simulation. The energy levels are split by a linear magnetic gradient field. Therefore, the internal states $\ket{\downarrow}$ and $\ket{\uparrow}$ of each site can be coupled by a local resonant microwave field.  (b) Spectroscopy measurement of the site occupation. At each frequency, we average the data over 5 repetitions, and integrate the signal along the $y-$axis. The blue circles are central positions of the atomic densities along $x$. According to the spatial position of the image, we plot the staircase structure of the lattice sites in cyan. The amplitude is normalized to the maximum atomic density. The sinusoidal fitting (orange-dashed line) shows the position of the sites and the staggered structure.}\label{Fig:Supp_SingleSite}
\end{figure}

We develop a technique to detect individual atoms at any specific site residing in the 1D optical lattice, without requiring a quantum gas microscope. By lifting the energy degeneracy of the transition frequency in each lattice site, we can flip the atomic state with a locally resonant microwave pulse. The potential used for shifting the energy levels is provided by a homogeneous magnetic gradient. We set the magnetic axis along the $x$ direction with a 7.3 G bias field, and meanwhile apply a $\sim$70 G/cm gradient field along this axis. In such a magnetic field, the energy level for the $\ket{\downarrow} \rightarrow \ket{\uparrow}$ transition is split by 5.6 kHz per lattice site. Addressing individual sites is realized by flipping the atomic internal level from $\ket{\downarrow}$ to $\ket{\uparrow}$ with a square $\pi-$pulse. Afterwards, the atom number on the corresponding site is recorded on a CCD camera with the $in\ situ$ absorption imaging.

This method enables the imaging of atoms with a spatial resolution better than the optical resolution of our imaging system. Instead, the resolution is determined by the energy splitting between lattice sites and the Fourier broadening of the microwave transition. For achieving such a high precision, we improve the stability of the magnetic field and position of the optical lattice. At an arbitrary microwave frequency, the position of the flipped stripe changes from shot-to-shot with a standard deviation of 0.11 $\mu$m. We set the Rabi frequency for the transition to 1.9 kHz, to make sure the Fourier broadening is smaller than the splitting between the neighboring sites. The number occupations on lattice sites are measured by scanning the microwave frequency. The frequency starts from 6.819104 GHz and ends at 6.819532 GHz, covering 75 sites of the optical lattice.

To benchmark our method, we perform this site-resolved imaging in a staggered state, as shown in Extended Data Fig. 2. Two essential features are captured by our measurement.
First, the detected atomic density oscillates with the same period as the site occupancy of the staggered state. Second, the central position of the flipped atoms follows the staircase behavior of the discrete lattice sites. We can clearly locate each individual lattice site in the 1D chain with this site-resolved imaging technique.

\subsection{Calibration of building block and reversibility of many-body dynamics}

Before performing the phase transition, the elementary parameters of the Hubbard model are calibrated precisely. The lattice depths are measured by applying a parametric excitation to the ground-band atoms. Then, we derive the Wannier functions of the atoms  at certain lattice depths, from which the on-site interaction $U$ and tunneling $J$ are obtained by integrating the overlap of Wannier functions. The linear potential $\Delta= 57$ Hz/site is formed by the projection of the gravity along the $x$-axis. The staggered offset $\delta$ is generated by the long lattice at a superlattice phase of $\varphi=\pi/2$.

\begin{figure}[!htb]
\includegraphics[width=7.9cm]{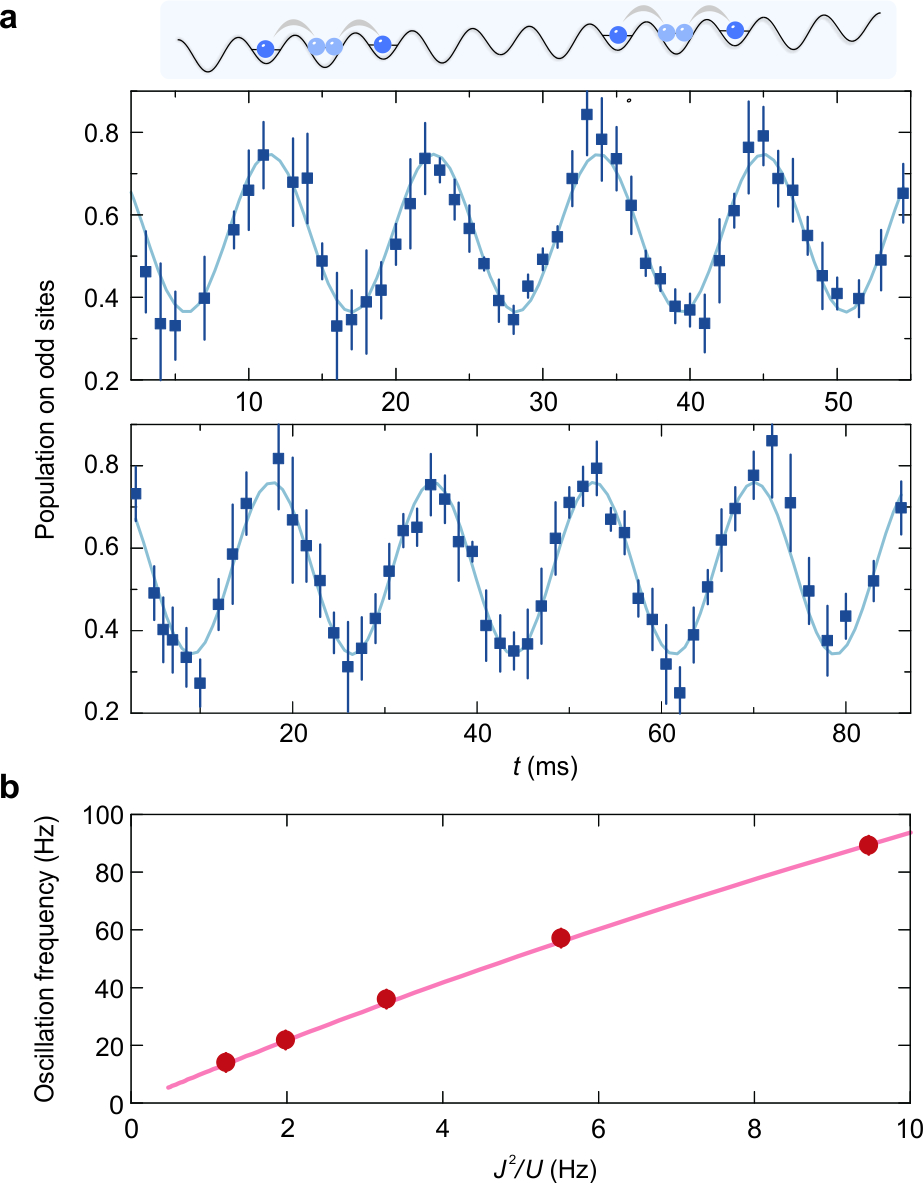}
\caption{Dynamics in building blocks. (a) We observe coherent evolutions in building blocks as sketched above the data. The solid curves are sinusoidal fitting results, which give oscillation frequencies of 89.3(3) Hz and 57.1(3) Hz, respectively. The error bars denote standard deviations. The small atom number in the dilute sample leads to larger statistical errors as compared to our many-body experiment reported in the main text. As these data show, the decay of oscillations is insignificant over a range of coupling strengths $\tilde t$, even for values larger than the greatest coupling strength used in the phase transition ($\tilde t=70$Hz). (b) The oscillation frequency and $J^2/U$ has an almost linear relation, which is in excellent quantitative agreement with the theoretical prediction based on the Bose-Hubbard model (solid curve).}
\label{Fig:fewtomany}
\end{figure}

We investigate the building block of our model and observe the coherent dynamics. To prepare a sample with isolated units, we quench the short lattice to $11.8(1)$ $E_r$ for $3$ ms starting from the staggered initial state. During such a short time, some of the atoms start to bind into doublons and enter the odd sites. We remove the majority atoms residing on the even sites, thereby creating a dilute sample with isolated building blocks. Afterwards, the superlattice is reshaped into the configuration of $\varphi=\pi/2$ and the dynamics of the atoms is monitored at the resonant condition with $U=1.17(1)$ kHz, $J = 105(1)$ Hz. The atoms oscillate between the state $\ket{020}$ and $\ket{101}$ via second-order hopping, forming an effective two-level system. In this building block, the self-energy correction shifts the resonant point to $U=2\delta - 4J^2/U$. Extended Data Fig. 3a shows a Rabi oscillation with negligible decay in this dilute sample, indicating an excellent coherence of the system. The amplitude of the oscillation is determined by the preparation fidelity of the building block.

\begin{figure}[!htb]
\includegraphics[width=7.5cm]{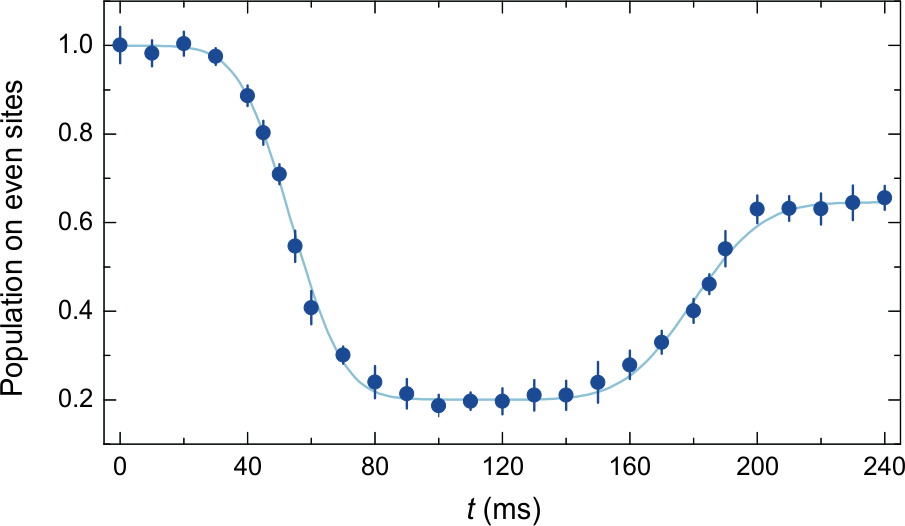}
\caption{Quantum phase transition and revival. In 240ms, we first ramp the mass from negative to positive. Then, we ramp the mass in the reversed direction, back to the symmetry-unbroken charge-proliferated phase. The recovery of the atoms on even sites indicates the reversibility of this phase transition. The solid curve is a guide for the eye.}
\label{Fig:revival}
\end{figure}

Another characteristic feature of a coherent adiabatic transition is its reversibility. Figure \ref{Fig:f2} shows a quantum phase transition from the charge-dominated phase to the C/P broken phase. In our sample with 71 sites, we compensate part of the residual potential of the blue-detuned lattice with a red-detuned dipole trap, which reduces the spatial inhomogeneity.
The coherence of the system allows us to recover the particle--anti-particle phase in another $120$ ms. We ramp the mass $m$ and tunneling $J$ in a reversed way as compared to the curves given in Fig.~\ref{Fig:f2}a, thereby decreasing $m/\tilde{t}$ from $11.6$ to $-39.8$ in order to return back to the charge-dominated phase. The occupancy of even sites is recovered to $0.66(3)$ in Extended Data Fig. 4, which is attributed mainly to the non-adiabaticity of the ramping process.

\subsection{Numerical calculations}

\begin{figure*}
\includegraphics[width=13.5cm]{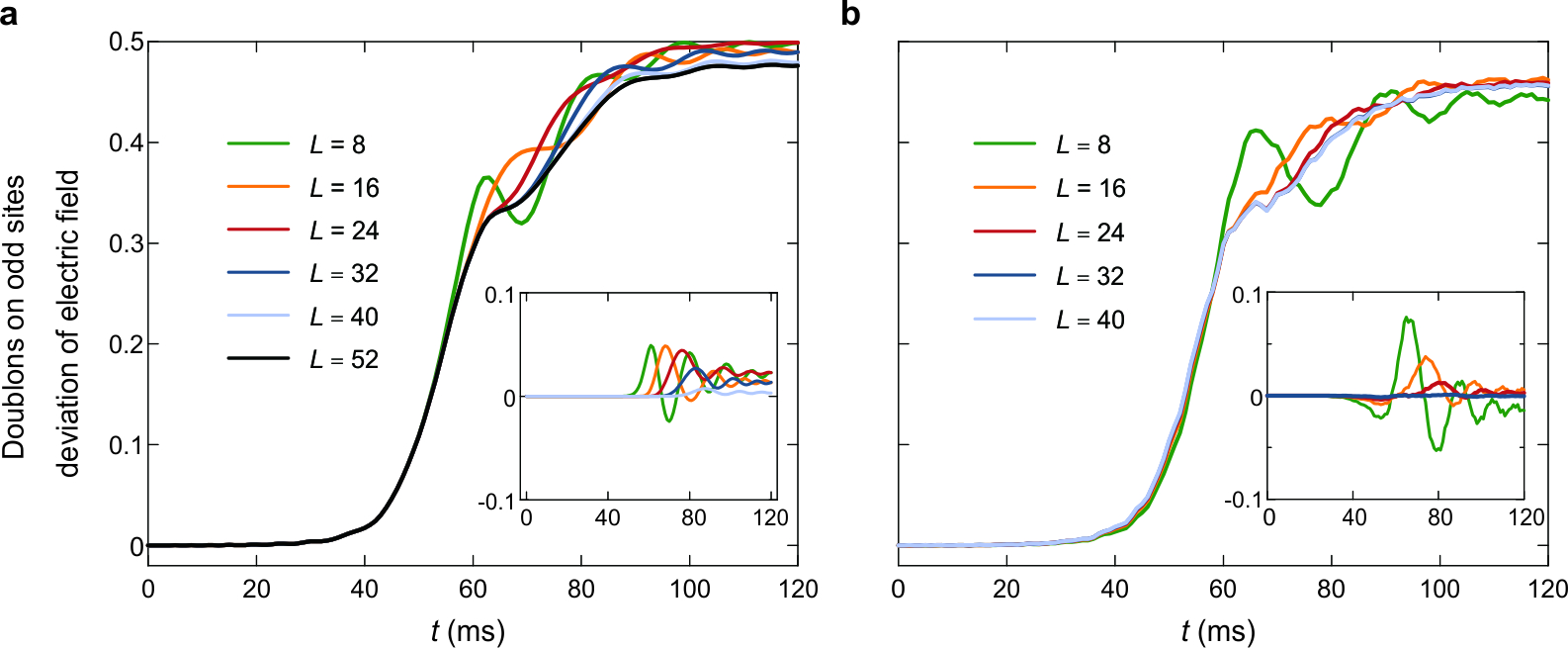}
\caption{Numerical simulations of the phase transition dynamics, calculated by (a) ED and (b) DMRG methods. We monitor the evolution of the deviation of the electric field, which corresponds to the double occupancy of the odd sites. (a) Simulations of the ideal, fully gauge-invariant QLM, using ED calculations under periodic boundary conditions. (b) Simulations of a 1D Bose--Hubbard system modeling our experiment, using the $t$-DMRG method.
The insets show the differences between results for different system sizes and the curve at the largest size ($L=52$ for ED and $L=40$ for $t$-DMRG), demonstrating finite-size convergence well below the range of experimental errors. }
\label{Fig:Convergence}
\end{figure*}

The dynamics of our $71$-site quantum simulator can be hardly computed up to the times we are interested in by classical numerical methods. However, we can calculate results at smaller system sizes and then check for convergence. To understand the quantum phase transition, we use exact diagonalization (ED) to calculate the QLM, and the time-adaptive density matrix renormalization group method ($t$-DMRG) to simulate the dynamics governed by the BHM.

To compute the dynamics in the ideal QLM, we use the mass $m$ and coupling strength $\tilde{t}$ as deduced from the Hubbard parameters. The time-dependent dynamics in the QLM fully obey Gauss's law. Using this conservation law to restrict to the implemented Hilbert space, we perform numerically exact diagonalizations for system sizes ranging from $L=8$ up to $L=52$ sites (see Extended Data Fig. 5a). Due to finite-size effects, the dynamics at smaller systems (such as $L=8$) show strong oscillations after crossing the critical point. We find that the non-adiabaticity caused by the ramping reduces the fidelity of our final state with increasing system size, due to the closure of the minimal gap at the critical point. The discrepancy between the curves for $L=40$ and $L=52$ is on the order of $10^{-3}$, indicating the volume convergence of our calculations. In Fig.~\ref{Fig:f4}b, the orange and blue curves for state populations are the ED results for system size of $L=52$.

We apply $t$-DMRG \cite{Schollwoeck:2011s,Mcculloch:2019s} to calculate the full dynamics of the 1D Bose-Hubbard chain. For our simulations, convergence is achieved at a time-step of $10^{-4}$s, truncation threshold of $10^{-6}$ per time-step, and maximum occupation of $2$ bosons per site. Finite-size effects are also investigated for several chain lengths. Similar to the behavior in the QLM, the dynamics becomes smooth with increasing chain length. Extended Data Fig. 5b shows volume convergence between the results for system sizes $L=32$ and $L=40$ sites. The theoretical predictions in Fig.~\ref{Fig:f2}c and Fig.~\ref{Fig:f4}b are obtained with system size $L=32$. Moreover, some imperfections of our system are taken into account in our $t$-DMRG calculations. Due to the inhomogeneity of the Gaussian-shaped $y$-lattice, the on-site interaction at the edge of the 71-site chain is about 10 Hz smaller than at the central site. Also, fluctuations of the depth of the long lattice lead to about $\pm$4.5 Hz uncertainty in the staggered energy $\delta$. Including these influences into our model, we estimate experimental observables with $\pm 1 \sigma$ confidence intervals (equal to the standard deviations).

These two numerical methods show consistent behavior at the converged system sizes, which means the discrepancies between our experiments and numerical calculations are not caused by finite-size effects. We attribute the remaining deviations to heating due to off-resonant excitations of the atoms by the optical lattice beams. Although the correlation length is $4.4^{+2.0}_{-1.0}$ sites and the domain size is 9(2) sites, both of the ED and $t$-DMRG calculations converge only once the system size is above about 40 sites, showing the essential role of many-body effects in the observed phenomena.

\subsection{Density-density correlations}

Constrained by the finite resolution of our microscope, we are not able to extract density-density correlations from the $in\ situ$ images. However, we can measure the correlation function by mapping the atomic distributions into momentum space. After a free expansion, the relation between the initial momentum $k_x$ and real-space position $x$ is $k_x = m_{\text{Rb}} x/t$. One characteristic momentum corresponding to the unity-filling Mott insulator is $k_x=2\hbar k$, which is related to the real-space position of $x_0=ht/(m_{\text{Rb}}\lambda/2)$. Then, the correlation function for a long chain with $N_{\text{ddc}}$ sites is,
\begin{equation}
C_k(x)  = 1 + \frac{1}{N_{\text{ddc}}^2} \sum_{i,j}e^{-\text{i} 2 \pi  x (i-j)/x_0} n_i n_j.
\label{eq:DensityCorrelation}
\end{equation}
Here, the position $x$ can be discretized into the pixels of the imaging plane. From this relation, we can easily find that the interference patterns emerge at a multiple of $x_0/d$, where $d$ is the periodicity of an ordered site occupation. Hence, the initial state with $d=2$ has first-order peaks at $k_x = \pm \hbar k$, and the states with $d=4$ would have peaks at $k_x = \pm 0.5 \hbar k$.

To detect the density-density correlations, we release the cloud and let it expand in the $xz$-plane for $8$ ms. The lattice depth along the $y$-axis is 25.6(2) $E_r$, which blocks the crosstalk of different 1D chains. Loosening the confinement along the $z$-axis strongly reduces the interaction between atoms, but also degrades the optical resolution. The characteristic length is $x_0 = 105\ \mu$m, which allows us to observe the new ordering with the microscope. We find that the initial size of the sample is comparatively smaller than the cloud after expansion. The exposure time for the absorption imaging is $10$ $\mu$s, thereby making the photon shot noise the major source of fluctuations on the signal.

The pattern in Fig.~\ref{Fig:f3}a is obtained by calculating the correlation function as defined in Eq.~\eqref{eq:DensityCorrelation}. For each image, the density correlation is the autocorrelation function. When we have a set of images, we perform this procedure using two different routes \cite{Folling:2005}. One is calculating the autocorrelation function for each image and then averaging them. Another is first averaging the images and then calculating the autocorrelation function once, which is used for normalizing the signal. Then we obtain the normalized density-density correlation. Such a method enables the extraction of correlations from noisy signals and is also robust to the cloud shape. The patterns in Fig.~\ref{Fig:f3}a are averaged over $523$ and $1729$ images, respectively. In the horizontal direction of the imaging plane, some stripes appear around $y=0$, which is caused by the fluctuations of the atomic center and total atom number \cite{Folling:2005}. Unlike the $in \ situ$ images, the atoms outside the region of interest still contribute to signals in momentum space.

\begin{figure}
\includegraphics[width=7.8cm]{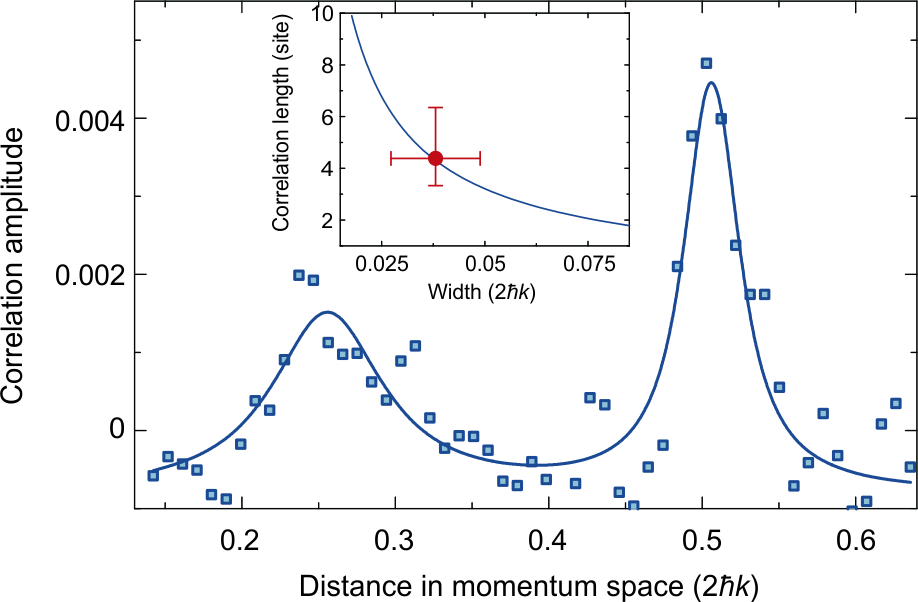}
\caption{Correlation length. We select the central region with two interfering peaks, where the background has been subtracted from the signal. The solid curve is a Lorentzian fitting of the data. The inset shows the relation between the peak width and the correlation length. The solid curve is calculated for a 1D system with $N_{\text{ddc}} = 100$ sites. The red point represents the correlation length of the final state as shown in Fig.~\ref{Fig:f3}a.}
\label{Fig:correlationlength}
\end{figure}

The correlation length is obtained from the width of the interference peak. For an entirely ordered state, such as the initial state, the amplitude of the density correlation is inversely proportional to the atom number, and the width is determined by the imaging resolution. However, spontaneous symmetry breaking in the phase transition induces the formation of domains.
At finite correlation length $\xi$, the peak width becomes broader. Assuming the correlation function decays exponentially in this 1D system, we can deduce $\xi$ from the peak width.
To extract the peak width, we first subtract the background profile from the correlation function. The background is a single-pixel section through the pattern center, whose direction is along the $-4^{\circ}$ with respect to the horizontal plane. As shown in Extended Data Fig. 6, we apply a Lorentzian fitting to the curve and find the width is $4.5 \pm 1.1 \ \mu$m.
The peaks at $\pm \hbar k$ and $\pm 2 \hbar k$ have widths of 2.0(2) $\mu$m and 1.9(4) $\mu$m respectively, which corresponds to the imaging resolution. Considering the broadening due to optical resolution, we obtain the correlation length as $\xi = 4.4^{+2.0}_{-1.0}$ sites.

\subsection{Potential violations of Gauss's law}

\begin{figure}
\includegraphics[width=7.7cm]{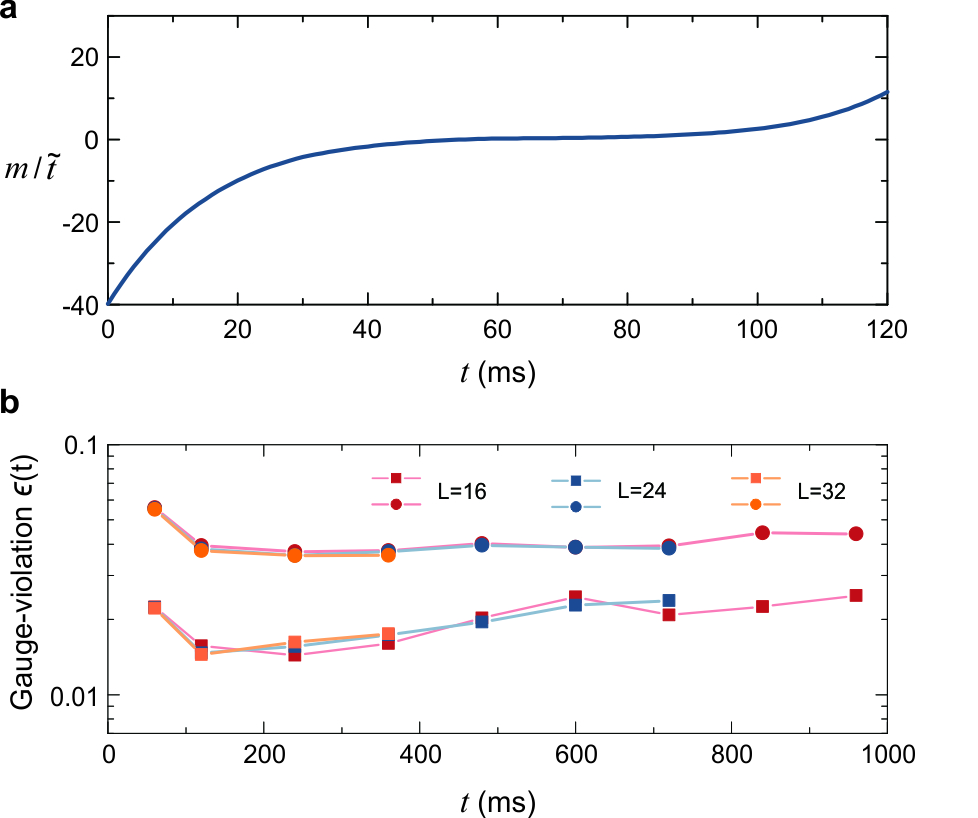}
\caption{Ramping speed and gauge violation. (a) The phase transition is driven by ramping the mass $m$ and the effective coupling $\tilde{t}$. We start from a large negative value of mass $m/\tilde{t}$, retain stronger coupling around the critical point, and end up with a large positive mass. (b) Gauge violation against total ramping time calculated with the $t$-DMRG method in a system with 16 (red), 24 (blue), and 32 (orange) optical-lattice sites. Using the same shape of the ramping curve in (a), we change the ramping speed by constraining the total ramping time. The squares points are the maxima of $\epsilon(t)$ throughout the dynamics, while the circles represent the gauge violations of the final states. Owing to the coherence of our many-body system, $\epsilon(t)$ reaches its maximum around the critical point, and decreases after crossing the critical point (see Fig. \ref{Fig:f4}b).}
\label{Fig:rampviolation}
\end{figure}

In our setup, potential gauge-violation terms arising from coherent processes are suppressed due to suitably engineered energy penalties. We can estimate the effect of these error terms in a three-site building block consisting of two--initially occupied--matter sites and the gauge link in between, described by the initial state $\ket{...n_{j} n_{j+1} n_{j+2}...}=\ket{...101...}$, $j$ even. The main cause of gauge violation stems from the desired matter--gauge-field coupling, which requires a second-order process of strength $\propto J^2/\delta$  involving the gauge-violating bare tunneling $J$. Similarly to a detuned Rabi oscillation, the population of the gauge-violating states $\ket{...110...}$ and $\ket{...011...}$ is on the order of $(J/\delta)^2$. At the highest coupling strength, which we reach at $t=60$ ms, we have $J/\delta = 0.13$, i.e., gauge-violating states have at most a few per cent of population. Rather than an incoherent dynamics that leads to accumulation of gauge violation over time, this bare tunneling is a coherent process that strongly mitigates the increasing of the induced gauge violation. In Fig. \ref{Fig:f4}b, the oscillations in the $t$-DMRG calculations are caused by such a detuned tunneling process. We further theoretically calculate the gauge violation of our system at long evolution time, as shown in Extended Data Fig. 7b. The gauge violation does not increase significantly even when the ramping time is about one order of magnitude longer than our experimental time scale.

At the same order of perturbation theory, direct tunneling between matter sites can occur, with a coupling strength on the order of $J^2/\delta$. This second-order tunneling is energetically suppressed by $U \pm 2\Delta$ when we consider the initial filling and the linear potential. The coherent oscillation in Fig.~\ref{Fig:fewtomany}a indicates that the atoms perform only the desired conversion between matter and gauge-field sites, and otherwise reside in their respective building blocks. Likewise, the staggered and linear potential suppress any long-range transport, which is also confirmed by the nearly constant size of the atomic cloud measured in absorption imaging.

If all gauge-violating many-body states experience such energy penalties, a deformed symmetry emerges that is perturbatively close to the ideal original one \cite{Chubb:2017s}, and which indefinitely suppresses gauge invariance-violating processes \cite{Halimeh:2020}. In the present case, the leading gauge-violating processes are energetically penalized, but violations at distant sites may in principle energetically compensate each other. This may lead to a slow leakage out of the gauge-invariant subspace through higher-order processes. On the experimentally relevant time scales, these processes are, however, irrelevant.

Our theoretical calculations, which are based on unitary time evolution, capture only coherent sources of gauge violations such as those mentioned above. As the agreement to our measured data suggests (see Fig.~\ref{Fig:f4}b), coherent processes are a main contribution to the weak gauge violation $\epsilon(t)$,  especially the first-order tunneling $J$.
In addition, there may appear dissipative processes that violate Gauss's law. Pure dephasing processes that couple to the atom density commute with Gauss's law and thus do not lead to gauge violations. In contrast, atom loss might affect gauge invariance. The lifetime characterizing the atom loss in optical lattices is $\sim$10 s, two orders magnitude longer than the duration of our sweep through the phase transition. Finally, in principle the lattice tilt introduces a finite lifetime of Wannier-Stark states \cite{Gluck:2002s}. Also this lifetime is much longer than experimentally relevant times. Our direct measurements of the violation of local gauge invariance corroborate the weakness of the various potential error sources over our experimental time scales.

\subsection{Measurement of Gauss's law violations}

\begin{figure}[!htb]
\includegraphics[width=7.6cm]{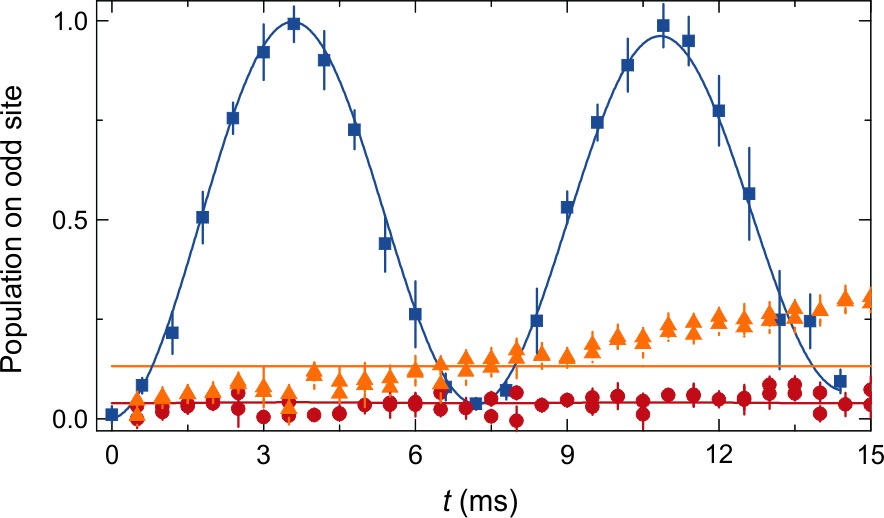}
\caption{Dynamics in double wells. Under the same superlattice configuration, we measure the evolution of three different states in DWs. The initial states are $\ket{10}$ (blue squares), $\ket{11}$ (orange triangles) and $\ket{20}$ (red circles), respectively. The state $\ket{10}$ oscillates with almost the full amplitude. The superexchange interaction drives the spin exchange process as expected. In contrast, the atom population remains constant for the state $\ket{20}$. The solid curves are exponentially damped sinusoidal fittings, where the frequency, phase, and decay rate are fixed. We find that the oscillation amplitude of tunneling is almost three orders of magnitude larger than the other two fitting values.}
\label{Fig:pairtunneling}
\end{figure}

\begin{figure*}
\includegraphics[width=15cm]{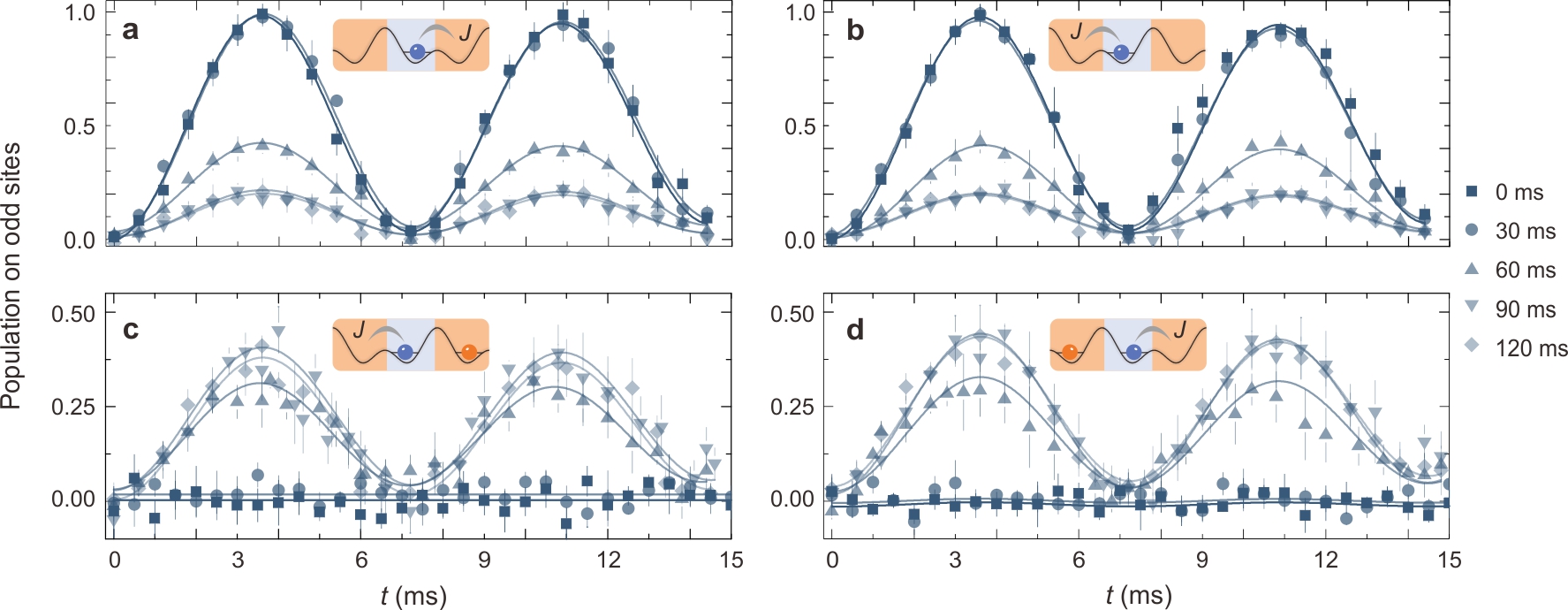}
\caption{Detecting the gauge-invariant states. We divide the atoms into DWs and then measure atom tunneling within each two-sites unit. (a)-(d) show the dynamics at four experimental sequences as sketched by the insets. Five different moments during the phase transition $t=0,\ 30,\ 60,\ 90,\ 120$ ms are selected for detecting the gauge-invariant states. We fit the data with a sinusoidal damping function, which has a period of 7.2 ms and an exponential decay constant of 96 ms. The amplitudes of the oscillations from panel (a) to (d) refer to $A^{(1)}_{\ket{10}}$, $A^{(1)}_{\ket{01}}$, $A^{(2)}_{\ket{10}}$ and $A^{(2)}_{\ket{01}}$, respectively. Then these amplitudes are used for calculating the state probabilities.}
\label{Fig:SAT}
\end{figure*}

\begin{figure*}
\includegraphics[width=13cm]{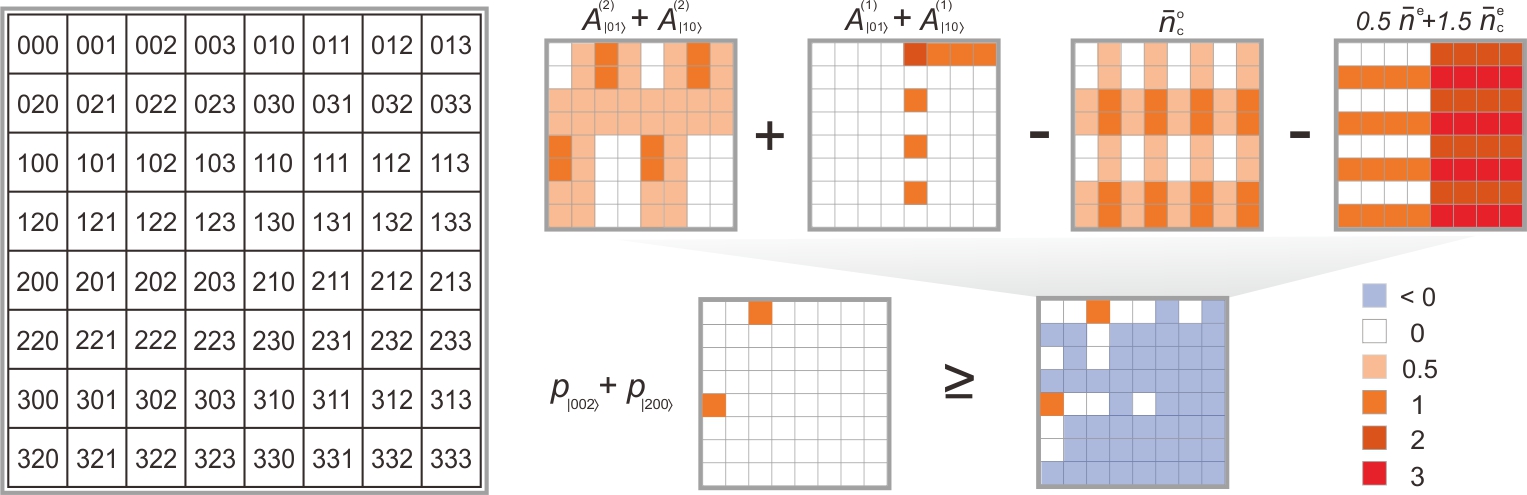}
\caption{Resolving the population of the states. For detecting the states $\ket{002}$ and $\ket{200}$, we extract their probabilities from several measurements. There are 64 states that may contribute to the oscillations, which are listed from $\ket{000}$ to $\ket{333}$ as a $8 \times 8$ square array. The amplitudes of these states according to our detection procedures are given by distinct colors. We use seven terms to deduce the lower bound for the probabilities as $p_{\ket{...002...}} + p_{\ket{...200...}} \geq A^{(2)}_{\ket{01}} + A^{(2)}_{\ket{10}} + A^{(1)}_{\ket{01}} + A^{(1)}_{\ket{10}} -\bar{n}^{\text{o}}_c - 0.5 \bar{n}^{\text{e}}-1.5 \bar{n}^{\text{e}}_c$. Such a relation can be captured from the checkerboard diagram.}
\label{Fig:StateMatrix}
\end{figure*}

We can verify the local fulfillment of Gauss's law---without the need for full state tomography---by measuring the probabilities of the gauge-invariant states. Considering the relevant three-site units, we can couple the central site with its left or right site by isolating the atoms into double wells. Then the sensing is achieved by the atom which can discriminate the filling number of its neighbor via subsequent dynamics. For the state $\ket{10}$ in a double well, the atom can tunnel and evolve to another state $\ket{01}$. The frequency is dramatically different from the state $\ket{20}$, which would tunnel in a pair with a strength of $4 J^2/U$. Since we mark the atoms with hyperfine levels, atoms in the $\ket{\downarrow\uparrow}$ would exchange their hyperfine state in a superexchange process. As shown in Extended Data Fig. 8, we observe the dynamics of these states by initially preparing them in double wells.
The superexchange frequency is $4 J^2/h U = 11$ Hz. However, the atom pairs cannot tunnel freely because this dynamics requires a further stabilization of the superlattice phase $\varphi$.
Even though the superexchange interaction can drive the evolution, such a process does not contribute to the oscillation amplitude at the frequency $2J/h$. In addition, the state $\ket{0n_j}$ does not contribute to the desired signal since we flip the hyperfine levels of the atoms on odd sites before implementing the tunneling sequence. We fit the oscillation with a function $y=y_0 + A e^{-t/\tau}\sin(2\pi f+\phi_0 )$. The frequency $f$, initial phase $\phi_0$, offset value $y_0$, and damping rate $\tau$ are fixed in the fitting. The signal is identified not only by the atom population but also by the characteristic frequency. Therefore, we can establish relations between the oscillation amplitude and the state probability.

Extended Data Fig. 9 shows the measurements for determining the population of gauge-invariant states. As illustrated in Fig.~\ref{Fig:f4}a, we monitor the oscillation of tunneling at four different sequences. After an evolution time $t$, the state detections begin with ramping the short lattice to 51.3(4) $E_r$. Then we tune the superlattice phase $\varphi$ from $\pi/2$ to 0 or $\pi$ and consequently divide the atoms into isolated double wells. In the procedure for detecting the state $\ket{010}$, we address and flip the hyperfine level of atoms residing the odd sites to $\ket{\uparrow}$, thereby mark the sites by their hyperfine levels. Afterwards, we quench the depth of the short- and long-lattice to 18.7(1) $E_r$ and 10.0(1) $E_r$ simultaneously. The atoms tunnel from even to odd sites within each double wells, whose expectation value is recorded by performing an absorption imaging. As shown in Extended Data Fig. 9a and b, the oscillation amplitudes almost equal to the ratios of even-site atoms.

For detecting the state $\ket{...002...}$ and $\ket{...200...}$, the procedure consists of more operations since the doublons cannot tunnel easily. Before the splitting of doublons, we remove the atoms residing on the even sites to ensure a 99.3(1)\% efficiency for the atom splitting. For instance, the state $\ket{12}$ in the DW would disturb the separation of doublons and also influence the following signal. Next, we perform the state flip operation and tune the superlattice phase from $0$ ($\pi$) to $\pi$ (0) to reach another configuration. In these double wells, the oscillations corresponding to atom tunneling are shown in Extended Data Fig. 9c and d. However, we should exclude the probability of other kinds of states, such as $\ket{...012...}$, since we remove the central particle and project it into $\ket{...002...}$. To clarify the process how we derive the final probability, the states that may contribute to the signals are listed  in a square array as Extended Data Fig. 10. Using seven experimental observable, we can extrapolate the population of the state $\ket{002}$ and $\ket{200}$. Actually, the other high-energy excitations, such as four particles per site, are also eliminated from this calculation. After performing the error propagation, the errors of the total probabilities are mainly arising from the shot noise of the absorption imaging. In Fig.~\ref{Fig:f4}b, the probabilities of the state $\ket{010}$ at $t=$ 0, 30 ms represent the gauge invariant terms with smaller errors.

Through these measurement, we are thus able to measure the probabilities of the states $\ket{...n_{j-1} n_{j} n_{j+1}...}=\ket{...010...}$, $\ket{...002...}$, and $\ket{...200...}$, $j$ even, from which we can compute the local projectors onto the gauge-invariant states, $P_\ell=\ket{010}\bra{010}+\ket{002}\bra{002}+\ket{200}\bra{200}$, where $\ell=j/2$ denotes the central matter site. These measurements enable us to certify the adherence to gauge invariance in our U(1) lattice-gauge quantum simulator.



\textbf{Acknowledgments} We thank J.~Berges, Q.~J.~Chen, Y.~J.~Deng, S.~Jochim, and W.~Zheng for discussions. We thank Z. Y. Zhou and G. X. Su for their help in the experimental measurements. This work is part of and supported by the DFG Collaborative Research Centre ``SFB 1225 (ISOQUANT)'',  the ERC Starting Grant StrEnQTh (Project-ID  804305), and the Provincia Autonoma di Trento.

\newpage
\onecolumngrid

\end{document}